\documentclass[11pt]{article}

\usepackage[a4paper,margin=1in]{geometry}
\usepackage{amsmath,amssymb,mathtools}
\usepackage{graphicx}
\usepackage{subcaption}
\usepackage{cite}
\usepackage{hyperref}
\usepackage{microtype}
\usepackage{enumitem}
\usepackage{bm}
\usepackage{booktabs}
\usepackage{tikz}
\usetikzlibrary{arrows.meta,calc,positioning,decorations.pathreplacing}
\usepackage{pgfplots}
\pgfplotsset{compat=1.18}
\usetikzlibrary{decorations.pathmorphing}
\usetikzlibrary{calc}
\usetikzlibrary{decorations.markings}
\usetikzlibrary{shapes.misc}
\usetikzlibrary{intersections, pgfplots.fillbetween}
\hypersetup{
  colorlinks=true,
  linkcolor=blue,
  citecolor=blue,
  urlcolor=blue
}

\newcommand{\JT}{\mathrm{JT}}
\newcommand{\CFT}{\mathrm{CFT}}

\newcommand{\St}{\widetilde S}
\newcommand{\Stgen}{\widetilde S^{\mathrm{gen}}}

\numberwithin{equation}{section}

\begin{document}

\vspace*{2.5cm}
\begin{center}
{\LARGE {Probing the Factorized Island Branch with the Capacity of Entanglement in JT Gravity} \\ \vspace*{1cm}}

\end{center}

\begin{center}
Ra\'ul Arias$^{\dagger}$\footnote{\texttt{rarias@fisica.unlp.edu.ar}},
Agust\'in Tamis$^{\dagger}$\footnote{\texttt{agustin.tamis@fisica.unlp.edu.ar}}
\end{center}

\begin{center}
{\footnotesize
\vspace{0.4cm}
$^\dagger$ Instituto de F\'isica La Plata -- CONICET and
Departamento de F\'isica, Universidad Nacional de La Plata, C.C.~67, 1900 La Plata, Argentina
}
\end{center}

\vspace*{0.5cm}

\vspace*{1.5cm}
\begin{abstract}
Black hole islands are usually diagnosed through the von Neumann entropy, but the full replica saddle contains more information than survives in the limit $n \to 1$. In this paper we show that the capacity of entanglement can detect that extra structure already within the controlled factorized island branch of JT gravity coupled to a large-$c$ bath. In the late-time high-temperature regime, the entropy plateau remains unchanged at the first nontrivial order, while the capacity acquires a definite correction. This provides a sharp semiclassical example in which nearby replica data are physically meaningful even when the entropy itself appears rigid. Our result shows that the factorized island saddle already carries finite-$n$ information beyond the entropy, and that the capacity is a natural observable for exposing it. More broadly, it highlights that the physics of island saddles is not exhausted by the $n=1$ limit: the surrounding replica geometry can contain additional, and observable, information about how the semiclassical saddle is assembled.
\end{abstract}

\newpage

\setcounter{tocdepth}{2}
\tableofcontents

\newpage

\section{Introduction}

The replica approach to black hole information has turned what once looked like a largely conceptual paradox into a concrete semiclassical problem of competing saddles. In this framework, generalized gravitational entropy, quantum extremal surfaces (QES), and finite-$n$ replica geometries are not merely formal devices: they are the objects that encode how fine-grained information is recovered from semiclassical gravity \cite{LewkowyczMaldacena2013,FaulknerLewkowyczMaldacena2013,EngelhardtWall2015,JafferisLewkowyczMaldacenaSuh2016,Dong2016}. In two-dimensional models, and especially in JT gravity coupled to nongravitating baths, this perspective has led to an analytically controlled understanding of islands, replica wormholes, and Page-curve physics \cite{AlmheiriEngelhardtMarolfMaxfield2019,AlmheiriMahajanMaldacenaZhao2020,Penington2020,PeningtonShenkerStanfordYang2022,AlmheiriHartmanMaldacenaShaghoulianTajdini2021,AlmheiriMahajanMaldacena2020}.

A central lesson of this progress is that the von Neumann entropy probes only one particularly special limit of a much richer replica problem. Away from $n=1$, the dominant saddle generally depends on the replica index in a nontrivial way, both through the generalized entropy functional and through the geometry that solves the replicated equations. This makes it natural to look for observables that are more sensitive than the entropy itself to nearby finite-$n$ structure. The \emph{capacity of entanglement} is a particularly sharp example. In quantum-information language it is the variance of the modular Hamiltonian, while in ``replica thermodynamics'' it plays the role of a heat capacity associated with the refined, or modular, entropy \cite{NakaguchiNishioka2016, deBoerJarvelaKeskiVakkuri2019}. In gravitational settings, it has already proved useful as a probe of replica wormholes and of the Page transition \cite{Kawabata2021,KawabataNishiokaOkuyamaWatanabe2021}.

At the same time, finite-$n$ calculations in JT gravity with baths remain technically nontrivial. The main difficulty is the conformal welding problem that determines the boundary map of the replicated saddle. The high-temperature, weak-backreaction regime
\begin{equation}
\kappa \equiv \frac{c\,\beta\,G_N}{6\pi \phi_r}\ll 1,
\label{eq:kappa-def}
\end{equation}
is already familiar from the original replica-wormhole analyses in JT gravity with baths \cite{Almheiri2020, HKP2024}. In the two-sided eternal black hole, they further isolated a late-time hierarchy
\begin{equation}
e^{-t_0}\ll \kappa \ll 1,
\qquad
t_0 \equiv \frac{2\pi}{\beta}\,t,
\label{eq:hierarchy-intro}
\end{equation}
for which the island branch factorizes, to the order they retained, into a pair of one-QES building blocks. This makes the setup especially well suited for asking controlled questions about finite-$n$ observables beyond the entropy.

The aim of the present paper is to isolate a specific physical question that arises naturally within this controlled framework: in the same late-time factorized high-temperature regime, does the island branch already carry finite-$n$ information that is invisible to the von Neumann entropy at $n=1$ but visible to the capacity of entanglement?

We answer this question in the affirmative. Working in the same dominant-mode truncation as in \cite{HKP2024}, we determine the first nontrivial $O(\kappa^2)$ correction to the generalized modular entropy on the factorized island branch in the late-time high-temperature regime. The key point is that this coefficient vanishes at $n=1$ while its replica derivative does not. As a consequence, the late-time island value of the von Neumann entropy remains unchanged at this order within the factorized approximation, whereas the corresponding late-time capacity plateau receives a definite negative shift proportional to $\kappa^2$.

This provides a concrete semiclassical example in which nearby replica data are physically meaningful even when the entropy itself appears rigid. In the present setup, the capacity is sensitive not only to the value of the island branch at $n=1$, but also to how that branch begins to vary away from $n=1$. The result therefore shows that the factorized island saddle already contains observable finite-$n$ information beyond what survives in the entropy limit.

At the same time, the scope of the claim is limited and clear. We are isolating the leading finite-$n$ effect intrinsic to the factorized island saddle itself, not solving the full two-sided replica problem beyond factorization. In particular, we do not determine the first genuinely non-factorizing inter-QES corrections. The point, however, is precisely that such effects are not needed in order to exhibit a clean physical distinction between entropy and capacity already within the controlled factorized branch.

The present work also fits into a growing literature on the capacity of entanglement as a diagnostic of modular structure. Besides the original discussions of its general properties and holographic interpretation \cite{NakaguchiNishioka2016,deBoerJarvelaKeskiVakkuri2019}, capacity has been used to study replica wormholes in JT gravity \cite{KawabataNishiokaOkuyamaWatanabe2021}, more general quantum-field-theoretic settings \cite{AriasDiGiulioKeskiVakkuriTonni2023}, and more recent finite-$n$ island constructions in JT and related two-dimensional dilaton models \cite{YuLinGe2026,AriasFondevila2026}. In this context, our goal is to isolate the leading local/factorized finite-$n$ effect on the JT island branch in the controlled high-temperature regime, and to compute it analytically.

The paper is organised as follows. In Section~\ref{sec:setup} we review the definitions of refined entropy and capacity and specify the regime of validity. Section~\ref{sec:welding} summarizes the perturbative ingredients from the high-temperature expansion that enter our calculation. Section~\ref{sec:main} contains the main computation of the $O(\kappa^2)$ island coefficient and derives the resulting shift of the capacity plateau. Appendix~\ref{app:equivalence} explains why differentiating the on-shell generalized modular entropy is equivalent to the more standard dilaton-derivative formula for the capacity. Appendix~\ref{app:nonfactorising} gives a parametric estimate of the first omitted non-factorizing correction. Appendix~\ref{app:m3mode} shows that the first omitted welding mode, $m=3$, only contributes at subleading order in the same factorized hierarchy. Finally, Appendix~\ref{app:numerics} collects numerical and internal consistency checks of the analytic result.

\section{Definitions and regime of validity}
\label{sec:setup}

In this section we briefly review the JT setup, fix our conventions for refined entropy and capacity, and state the regime in which the computation performed below is controlled. Our discussion follows the same physical setup as in \cite{HKP2024}: an eternal JT black hole coupled to large-$c$ nongravitating baths, in a perturbative high-temperature regime where the replica geometry can be treated analytically.

\subsection{JT gravity coupled to nongravitating baths}
\label{subsec:JT-baths}

We consider Jackiw--Teitelboim gravity in two dimensions coupled to a conformal matter sector. In Euclidean signature, and setting the AdS$_2$ radius to one, the action may be written as
\begin{equation}
\begin{aligned}
I_{\JT}
&=
-\frac{S_0}{2\pi}
\left[
\frac12\int_{\mathcal M}d^2x\,\sqrt{g}\,R
+
\int_{\partial\mathcal M}d u\,\sqrt{\gamma}\,K
\right]
-\frac{1}{16\pi G_N}\int_{\mathcal M}d^2x\,\sqrt{g}\,\Phi\,(R+2)
\\
&\quad
-\frac{1}{8\pi G_N}\int_{\partial\mathcal M}d u\,\sqrt{\gamma}\,\Phi\,(K-1)
+
I_{\CFT}[g,\chi].
\end{aligned}
\label{eq:JT-action}
\end{equation}
Here $S_0$ is the extremal entropy, $\Phi$ is the dilaton, and $I_{\CFT}$ denotes the action of a large-$c$ conformal field theory coupled to the metric. The first term is topological and controls the genus weighting of replica saddles, while the second term imposes constant negative curvature and supplies the dynamical dilaton contribution \cite{Teitelboim1983,Jackiw1984,AlmheiriPolchinski2015,EngelsoyMertensVerlinde2016,MaldacenaStanfordYang2016}. Near the asymptotic boundary, the dilaton obeys the standard JT boundary condition
\begin{equation}
\Phi|_{\partial \mathcal M}\sim \phi_r/\epsilon ,
\end{equation}
with $\phi_r$ the renormalized boundary value that controls the Schwarzian limit.

The Lorentzian geometry corresponding to this Euclidean saddle is the usual two-sided eternal
AdS$_2$ black hole at inverse temperature $\beta$. The Lorentzian continuation of \eqref{eq:JT-action} contains the standard two-sided JT black hole coupled,
in the bath setup of interest here, to two flat nongravitating half-lines through transparent
interfaces. Since the computation below is naturally formulated in the Euclidean replica geometry,
it is more convenient to introduce directly the disk description that will be used in Section~\ref{sec:welding}.

In Euclidean signature the gravitational region is represented by the unit disk,
\begin{equation}
ds^2 = \frac{4\,dw\,d\bar w}{(1-|w|^2)^2}, \qquad |w|<1 ,
\label{eq:diskmetric}
\end{equation}
with the physical boundary described by a reparametrized unit circle
\begin{equation}
w(\tau)=e^{i\theta(\tau)}, \qquad \tau\sim \tau+2\pi .
\label{eq:boundarycurveintro}
\end{equation}
In the one-QES building block, the finite-$n$ quantum extremal surface is located at a point
$a$ in the unit disk, and the corresponding replica geometry is encoded by the welding maps
introduced later. This is the local description that controls the factorized late-time island branch.

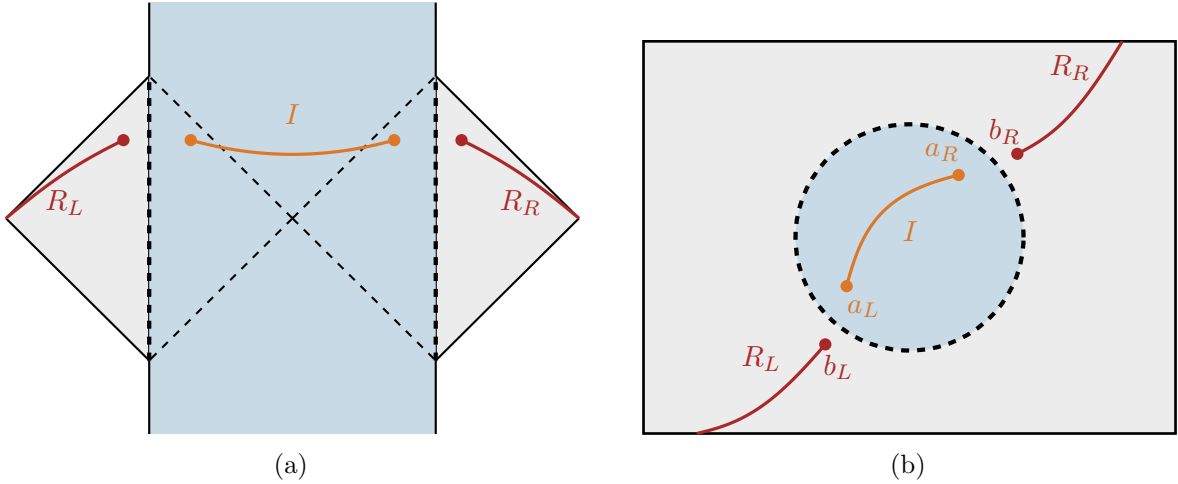
\begin{figure}[t]
\centering

% ===================== (a) =====================
\begin{subfigure}[t]{0.49\textwidth}
\centering
\begin{tikzpicture}[scale=2.6]

\definecolor{celestesave}{RGB}{200,218,230}
\definecolor{puntito}{RGB}{220,120,40}
\definecolor{rojito}{RGB}{170,40,40}

% Tamaño de puntos (DIÁMETRO) en unidades absolutas (NO escala)
\def\ptDia{4.6pt}

% Parámetros geométricos
\def\xmax{1.45}
\def\ymin{-0.4}
\def\ymax{1.8}

\pgfmathsetmacro{\yc}{(\ymin+\ymax)/2}
\pgfmathsetmacro{\xmid}{\xmax/2}
\pgfmathsetmacro{\Delta}{\xmax-\xmid}

% Rectángulo
\fill[celestesave] (0,\ymin) rectangle (\xmax,\ymax);

% Triángulos externos (derecha)
\fill[gray!15]
  ({\xmax+\Delta},\yc)
  -- (\xmax,{\yc+\Delta})
  -- (\xmax,{\yc-\Delta})
  -- cycle;

\draw[thick] ({\xmax+\Delta},\yc) -- (\xmax,{\yc+\Delta});
\draw[thick] ({\xmax+\Delta},\yc) -- (\xmax,{\yc-\Delta});
\draw[thick] (\xmax,{\yc+\Delta}) -- (\xmax,{\yc-\Delta});

% Triángulos externos (izquierda)
\fill[gray!15]
  ({-\xmid},\yc)
  -- (0,{\yc+\Delta})
  -- (0,{\yc-\Delta})
  -- cycle;

\draw[thick] ({-\xmid},\yc) -- (0,{\yc+\Delta});
\draw[thick] ({-\xmid},\yc) -- (0,{\yc-\Delta});
\draw[thick] (0,{\yc+\Delta}) -- (0,{\yc-\Delta});

% Triángulos internos (derecha)
\fill[celestesave]
  (\xmid,\yc)
  -- (\xmax,{\yc+\Delta})
  -- (\xmax,{\yc-\Delta})
  -- cycle;

\draw[thick,dashed] (\xmid,\yc) -- (\xmax,{\yc+\Delta});
\draw[thick,dashed] (\xmid,\yc) -- (\xmax,{\yc-\Delta});

% Triángulos internos (izquierda)
\fill[celestesave]
  (\xmid,\yc)
  -- (0,{\yc+\Delta})
  -- (0,{\yc-\Delta})
  -- cycle;

\draw[thick,dashed] (\xmid,\yc) -- (0,{\yc+\Delta});
\draw[thick,dashed] (\xmid,\yc) -- (0,{\yc-\Delta});

% Bordes rectángulo
\draw[thick] (0,\ymin) -- (0,{\yc-\Delta});
\draw[thick] (0,{\yc+\Delta}) -- (0,\ymax);
\draw[thick] (\xmax,\ymin) -- (\xmax,{\yc-\Delta});
\draw[thick] (\xmax,{\yc+\Delta}) -- (\xmax,\ymax);

% Separación dashed (las que cruzan el rectángulo)
\draw[line width=1.6pt,dashed] (0,{\yc-\Delta}) -- (0,{\yc+\Delta});
\draw[line width=1.6pt,dashed] (\xmax,{\yc-\Delta}) -- (\xmax,{\yc+\Delta});

% =====================================================
% Puntitos rojos (exteriores) + CURVAS rojas
% =====================================================
\pgfmathsetmacro{\yDotR}{\yc+0.55*\Delta}
\pgfmathsetmacro{\xDotOffR}{0.18*\Delta}

% coordenadas de puntas exteriores
\pgfmathsetmacro{\xL}{-\xmid}
\pgfmathsetmacro{\yL}{\yc}
\pgfmathsetmacro{\xE}{\xmax+\Delta}
\pgfmathsetmacro{\yE}{\yc}

% coordenadas de puntos rojos
\pgfmathsetmacro{\xRleft}{-\xDotOffR}
\pgfmathsetmacro{\yRleft}{\yDotR}
\pgfmathsetmacro{\xRright}{\xmax+\xDotOffR}
\pgfmathsetmacro{\yRright}{\yDotR}

% Curvas rojas: media ondulación + concavidad hacia arriba
\pgfmathsetmacro{\Are}{0.02}
\draw[rojito, line width=1.2pt]
  plot[domain=0:1, samples=220, variable=\s]
  ({\xL + (\xRleft-\xL)*\s},
   {\yL + (\yRleft-\yL)*\s + 0.04*\s*(1-\s) + \Are*sin(180*\s)});

\draw[rojito, line width=1.2pt]
  plot[domain=0:1, samples=220, variable=\s]
  ({\xE + (\xRright-\xE)*\s},
   {\yE + (\yRright-\yE)*\s + 0.04*\s*(1-\s) + \Are*sin(180*\s)});

% ---- Etiquetas R (abajo de las curvas rojas) ----
\node[rojito, font=\bfseries]
  at ({(\xL+\xRleft)/2},{(\yL+\yRleft)/2 - 0.12}) {$R_L$};

\node[rojito, font=\bfseries]
  at ({(\xE+\xRright)/2},{(\yE+\yRright)/2 - 0.12}) {$R_R$};

% Puntos rojos (NO escalan)
\node[circle, fill=rojito, inner sep=0pt, minimum size=\ptDia, transform shape=false]
  at (\xRleft,\yRleft) {};
\node[circle, fill=rojito, inner sep=0pt, minimum size=\ptDia, transform shape=false]
  at (\xRright,\yRright) {};

% =====================================================
% Puntitos naranjas (interiores) + CURVA naranja
% ahora a la MISMA altura que los rojos
% =====================================================

% Misma altura que los rojos
\pgfmathsetmacro{\yDotI}{\yDotR}

% A esta altura, la diagonal superior izquierda va de (xmid,yc) a (0,yc+Delta)
% t=(y-yc)/Delta, x_on_diag = xmid*(1-t)
\pgfmathsetmacro{\tI}{(\yDotI-\yc)/\Delta}
\pgfmathsetmacro{\xOnDiagL}{\xmid*(1-\tI)}

% Para no tocar la frontera (dejo un margen)
\pgfmathsetmacro{\shrink}{0.08*\xOnDiagL}
\pgfmathsetmacro{\xOnDiagLsafe}{\xOnDiagL-\shrink}

% Los querés cerca de la frontera interior: factor 0.7 (tu elección)
\pgfmathsetmacro{\xDotI}{0.7*\xOnDiagLsafe}

% coordenadas naranjas
\pgfmathsetmacro{\xPleft}{\xDotI}
\pgfmathsetmacro{\yPleft}{\yDotI}
\pgfmathsetmacro{\xPright}{\xmax-\xDotI}
\pgfmathsetmacro{\yPright}{\yDotI}

% Curva naranja: concavidad hacia ABAJO (∩ clara y limpia)
\draw[puntito, line width=1.2pt]
  plot[domain=0:1, samples=200, variable=\s]
  ({\xPleft + (\xPright-\xPleft)*\s},
   {\yPleft - 0.4*\Delta*\s*(1-\s)});

% ---- Etiqueta I (arriba de la curva naranja) ----
\node[puntito, font=\bfseries]
  at ({(\xPleft+\xPright)/2},{\yPleft + 0.18*\Delta}) {$I$};

% Puntos naranjas (NO escalan)
\node[circle, fill=puntito, inner sep=0pt, minimum size=\ptDia, transform shape=false]
  at (\xPleft,\yPleft) {};
\node[circle, fill=puntito, inner sep=0pt, minimum size=\ptDia, transform shape=false]
  at (\xPright,\yPright) {};

\end{tikzpicture}
\caption{}
\label{fig:xxx-a}
\end{subfigure}
\hfill
% ===================== (b) =====================
\begin{subfigure}[t]{0.49\textwidth}
\centering
\begin{tikzpicture}

\definecolor{celestesave}{RGB}{200,218,230}
\definecolor{puntito}{RGB}{220,120,40}
\definecolor{rojito}{RGB}{170,40,40}

% Tamaño de puntos (DIÁMETRO) en unidades absolutas (NO escala)
\def\ptDia{4.6pt}

% Ancho de curvas
\def\curveLW{1.2pt}

\def\R{1.5cm}
\def\W{7.0cm}
\def\H{5.2cm}

\fill[gray!15] (-\W/2,-\H/2) rectangle (\W/2,\H/2);

\draw[line width=1pt] (-\W/2,-\H/2) rectangle (\W/2,\H/2);

\filldraw[
  fill=celestesave,
  draw=black,
  line width=1.6pt,
  dashed
] (0,0) circle (\R);

% -------------------------------------------------
% Separaciones a la frontera
% -------------------------------------------------

\pgfmathsetlengthmacro{\gap}{0.30cm}
\pgfmathsetlengthmacro{\gapIn}{0.45cm}

\pgfmathsetlengthmacro{\rOut}{\R + \gap}
\pgfmathsetlengthmacro{\rIn}{\R - \gapIn}

% -------------------------------------------------
% Ángulos (variación suave)
% -------------------------------------------------

\def\angRedR{38}
\def\angOraR{52}

\def\angOraL{218}
\def\angRedL{232}

% Coordenadas polares (radio fijo)
\coordinate (dotR_L) at ({\rOut*cos(\angRedL)},{\rOut*sin(\angRedL)});
\coordinate (dotR_R) at ({\rOut*cos(\angRedR)},{\rOut*sin(\angRedR)});
\coordinate (dotO_L) at ({\rIn*cos(\angOraL)},{\rIn*sin(\angOraL)});
\coordinate (dotO_R) at ({\rIn*cos(\angOraR)},{\rIn*sin(\angOraR)});

% -------------------------------------------------
% Puntos
% -------------------------------------------------

\node[circle, fill=rojito, inner sep=0pt, minimum size=\ptDia, transform shape=false] at (dotR_L) {};
\node[circle, fill=rojito, inner sep=0pt, minimum size=\ptDia, transform shape=false] at (dotR_R) {};

\node[circle, fill=puntito, inner sep=0pt, minimum size=\ptDia, transform shape=false] at (dotO_L) {};
\node[circle, fill=puntito, inner sep=0pt, minimum size=\ptDia, transform shape=false] at (dotO_R) {};

% Etiquetas de los puntitos
\node[rojito, font=\bfseries] at ($(dotR_L)+(0.17,-0.3cm)$) {$b_L$};
\node[rojito, font=\bfseries] at ($(dotR_R)+(-0.17,0.3cm)$) {$b_R$};

\node[puntito, font=\bfseries] at ($(dotO_L)+(0.22cm,-0.3cm)$) {$a_L$};
\node[puntito, font=\bfseries] at ($(dotO_R)+(-0.22cm,0.3cm)$) {$a_R$};

% -------------------------------------------------
% Curva naranja simétrica
% -------------------------------------------------

\pgfmathsetlengthmacro{\Aora}{0.24cm}
\pgfmathsetlengthmacro{\Bora}{0.48cm}

\path (dotO_L); \pgfgetlastxy{\xOc}{\yOc}
\path (dotO_R); \pgfgetlastxy{\xOd}{\yOd}

\pgfmathsetmacro{\dxO}{\xOd-\xOc}
\pgfmathsetmacro{\dyO}{\yOd-\yOc}
\pgfmathsetmacro{\lenO}{veclen(\dxO,\dyO)}
\pgfmathsetmacro{\nxO}{-(\dyO/\lenO)}
\pgfmathsetmacro{\nyO}{ (\dxO/\lenO)}

\pgfmathsetmacro{\sgnO}{ifthenelse(\nyO<0,-1,1)}
\pgfmathsetmacro{\nxO}{\sgnO*\nxO}
\pgfmathsetmacro{\nyO}{\sgnO*\nyO}

\draw[puntito, line width=\curveLW]
  plot[domain=0:1, samples=240, variable=\s]
  ({\xOc + \dxO*\s + \nxO*(\Bora*\s*(1-\s) + \Aora*sin(180*\s))},
   {\yOc + \dyO*\s + \nyO*(\Bora*\s*(1-\s) + \Aora*sin(180*\s))});

\node[puntito, font=\bfseries]
  at ($ (dotO_L)!0.5!(dotO_R) + (0.1cm,0cm) $) {$I$};

% -------------------------------------------------
% Puntos de arranque en la frontera del rectángulo
% -------------------------------------------------

\coordinate (rectBL) at (-\W/2,-\H/2);
\coordinate (rectTR) at (\W/2,\H/2);

% MISMA separación 0.70cm
\coordinate (startL) at ($ (rectBL) + (0.70cm,0) $);  % ahora hacia la derecha
\coordinate (startR) at ($ (rectTR) + (-0.70cm,0) $); % hacia la izquierda

% -------------------------------------------------
% Curvas rojas (cóncavas hacia abajo)
% -------------------------------------------------

\pgfmathsetlengthmacro{\Ared}{0.18cm}
\pgfmathsetlengthmacro{\Bred}{0.22cm}

\path (startL); \pgfgetlastxy{\xSa}{\ySa}
\path (dotR_L); \pgfgetlastxy{\xEa}{\yEa}

\draw[rojito, line width=\curveLW]
  plot[domain=0:1, samples=220, variable=\s]
  ({\xSa + (\xEa-\xSa)*\s},
   {\ySa + (\yEa-\ySa)*\s - \Bred*\s*(1-\s) - \Ared*sin(180*\s)});

\path (startR); \pgfgetlastxy{\xSb}{\ySb}
\path (dotR_R); \pgfgetlastxy{\xEb}{\yEb}

\draw[rojito, line width=\curveLW]
  plot[domain=0:1, samples=220, variable=\s]
  ({\xSb + (\xEb-\xSb)*\s},
   {\ySb + (\yEb-\ySb)*\s - \Bred*\s*(1-\s) - \Ared*sin(180*\s)});

% ---- Etiquetas R (arriba) ----
\node[rojito, font=\bfseries]
  at ($ (startL)!0.5!(dotR_L) + (0,0.4cm) $) {$R_L$};

\node[rojito, font=\bfseries]
  at ($ (startR)!0.5!(dotR_R) + (0,0.4cm) $) {$R_R$};

\end{tikzpicture}
\caption{}
\label{fig:xxx-b}
\end{subfigure}

\caption{Two-sided setup for the entropy of two bath intervals in JT gravity coupled to
non-gravitating baths. (a) Lorentzian constant-time picture, with radiation region
$R=R_L\cup R_R$, bath endpoints $b_L,b_R$, and QES endpoints $a_L,a_R$. The
corresponding island $I$ lies inside the gravitating region. (b) Euclidean disk
representation of the same endpoint data.}
\label{fig:two-intervals-setup}
\end{figure}

The radiation region whose entropy we study is the union
\begin{equation}
R = R_L \cup R_R
\label{eq:Rdef}
\end{equation}
of two bath intervals, one in each asymptotic region, anchored at equal boundary times. A schematic picture is shown in Fig.~\ref{fig:two-intervals-setup}. On the
Hawking branch, the entropy is computed without an island. On the island branch, the relevant
generalized entropy is extremized with respect to one QES on each side, so that the island is the
union of two intervals inside the gravitating region. In the regime of interest below, this two-sided
problem simplifies substantially and effectively reduces to two identical one-QES building blocks.

\subsection{Refined entropy, generalized modular entropy, and capacity}
\label{subsec:refined-capacity}

We now recall the finite-$n$ observables that will be used throughout the paper. Let $Z(n)$ denote the replica partition function evaluated on a chosen saddle branch. Following \cite{NakaguchiNishioka2016,deBoerJarvelaKeskiVakkuri2019,KawabataNishiokaOkuyamaWatanabe2021}, we define
\begin{equation}
{\cal{F}}(n)\equiv -\frac{1}{n}\log Z(n),
\qquad
\St(n)\equiv n^2\partial_n {\cal{F}}(n).
\label{eq:refined-entropy-def}
\end{equation}
The quantity $\St(n)$ is the \emph{refined entropy}, also called modular entropy. It is the natural finite-$n$ object associated with the replica family, and reduces to the von Neumann entropy at $n=1$.

The \emph{capacity of entanglement} is then defined by
\begin{equation}
C^{(n)}\equiv -n\,\partial_n \St(n),
\qquad
C\equiv C^{(n)}\big|_{n=1}.
\label{eq:capacity-def}
\end{equation}
Equivalently, $C$ is the variance of the modular Hamiltonian in the state under consideration. In the present gravitational context it is crucial that the derivative in \eqref{eq:capacity-def} is taken \emph{within a fixed saddle branch}, and only afterwards evaluated at $n=1$. In particular, one does not extremize the capacity itself.

In gravitational problems with islands, the relevant quantity is the \emph{generalized} refined entropy. At fixed replica index $n$, this is obtained by evaluating the dilaton contribution and the matter refined entropy on the $n$-dependent QES:
\begin{equation}
\Stgen_n
=
\frac{\Phi^{(n)}_{\rm QES}}{4G_N}
+
\St_n^{\CFT}.
\label{eq:Stgen-def}
\end{equation}
For the one-QES building block, \eqref{eq:Stgen-def} is the object to be extremized with respect to the QES position. In the full two-sided island saddle one must in addition restore the topological contribution proportional to $S_0$, which we will do explicitly when writing the complete branch formula in Section~\ref{sec:main}. For the Hawking branch there is no QES extremization, whereas for the island branch the extremization of \eqref{eq:Stgen-def} determines a finite-$n$ replica surface whose location depends nontrivially on $n$ \cite{Dong2016,HKP2024}. 

This distinction is precisely why the capacity is interesting here. The entropy probes only the $n\to 1$ limit of the branch, while the capacity is sensitive to the first nontrivial variation of the finite-$n$ saddle around that limit. Our goal is to isolate this information in the controlled late-time factorized regime of the JT island branch.

\subsection{High-temperature regime and factorization hierarchy}
\label{subsec:regime-validity}

The perturbative parameter that organizes the finite-$n$ expansion is
\begin{equation}
\kappa
=
\frac{c\,\beta\,G_N}{6\pi \phi_r}
\ll 1.
\label{eq:kappa-regime}
\end{equation}
Physically, $\kappa$ measures the strength of the matter backreaction relative to the dilaton slope at the asymptotic boundary. It controls the size of the boundary reparametrization and makes the conformal welding problem perturbative \cite{Almheiri2020,HKP2024}.

In the eternal black hole setup we further work in the late-time regime 
\begin{equation}
e^{-t_0}\ll \kappa\ll 1,
\qquad
t_0=\frac{2\pi t}{\beta},
\label{eq:factorized-regime}
\end{equation}
In this regime the genuinely two-sided part of the replica geometry is parametrically suppressed, and the island saddle factorizes, to the order kept here, into a sum of two identical one-sided contributions. This is the approximation under which the two-sided problem becomes analytically manageable.

It is worth being explicit about what this means. The original one-sided building block is naturally organized as a large-$|b|$ expansion, where the bath endpoint parameter $b$ controls the late-time hierarchy and the physical boundary modes scale as $b^{-m}$ for $m>1$.
Since the modes $m=0,\pm1$ are pure $\mathrm{PSL}(2,\mathbb R)$ gauge, the first physical mode is $m=2$, which is therefore the least suppressed contribution in the factorized regime. This is the origin of the dominant-mode truncation adopted in \cite{HKP2024} and also used in the present paper.

The final one-sided formulas are later evaluated on the representative $|b|=1$ building block, but the logic is important: the truncation is justified in the original large-$|b|$ factorized regime, and only afterwards is the resulting truncated one-sided answer evaluated at $|b|=1$. In particular, the first omitted mode, $m=3$, does not compete with the leading $m=2$ contribution; rather, it enters with an additional suppression in the same large-$|b|$ hierarchy. We make this explicit in Appendix~\ref{app:m3mode}, where we show that it contributes only to the first subleading factorized correction.

Accordingly, every result in this paper should be read with the following qualifications in mind. First, we work within the same dominant-mode truncation as \cite{HKP2024}. Second, we do not include the genuinely non-factorizing inter-QES corrections, which are suppressed by powers of $e^{-2t_0}$ in the regime \eqref{eq:factorized-regime}. Our main result should therefore be interpreted as the leading finite-$n$ correction intrinsic to the \emph{factorized} island branch, not as the complete answer to the fully global two-sided replica problem.

These restrictions are also what make the calculation clean. Once one stays inside the regime \eqref{eq:kappa-regime}--\eqref{eq:factorized-regime}, the first nontrivial finite-$n$ correction to the island-branch capacity can be extracted analytically, and its physical interpretation is unambiguous.

\section{Perturbative boundary dynamics and one-sided replica data}
\label{sec:welding}

In this section we collect the ingredients from the high-temperature analysis of
\cite{HKP2024} that enter our computation. The goal is not to repeat the full derivation,
but rather to make explicit which data of the finite-$n$ replica saddle are needed at each
step, and why the factorized late-time island branch reduces to a controlled one-QES problem.

The discussion naturally splits into two parts. First, one has the boundary dynamics of the
replicated JT saddle, encoded in a reparametrization mode \(\theta(\tau)\) determined by a
boundary equation of motion. Second, once that boundary curve is known, the matter refined
entropy is expressed in terms of the corresponding welding maps \(F\) and \(G\), together
with the finite-$n$ QES position \(a\). In the factorized regime of interest, only the first
few pieces of this data are needed.

\subsection{Boundary equation of motion}

The replicated Euclidean boundary is described by a map
\begin{equation}
w(\tau)=e^{i\theta(\tau)}, \qquad \tau\sim \tau+2\pi ,
\label{12}
\end{equation}
where $\tau$ is the dimensionless angular Euclidean time,
\begin{equation}
\tau=\frac{2\pi}{\beta}\,\tau_{\rm phys},
\qquad
\tau_{\rm phys}\sim \tau_{\rm phys}+\beta .
\end{equation}
Here $\tau_{\rm phys}$ is the physical Euclidean time along the asymptotic thermal boundary. In
terms of $\tau$, the JT boundary equation is
\begin{equation}
\frac{\phi_r}{4\beta G_N}\,\partial_\tau B(\tau)
=
i\Big(T_{yy}(i\tau)-T_{\bar y\bar y}(-i\tau)\Big),
\qquad
B(\tau)\equiv \{w,\tau\}+2\,T(w)\,\bigl(w'(\tau)\bigr)^2,
\label{eq:boundary-eom}
\end{equation}
where primes denote derivatives with respect to \(\tau\), \(\{w,\tau\}\) is the Schwarzian,
and
\begin{equation}
T(w)=\frac12\{W,w\}
\label{eq:Liouville-stress}
\end{equation}
is the Liouville stress tensor associated with the replica uniformization map \(W(w)\).

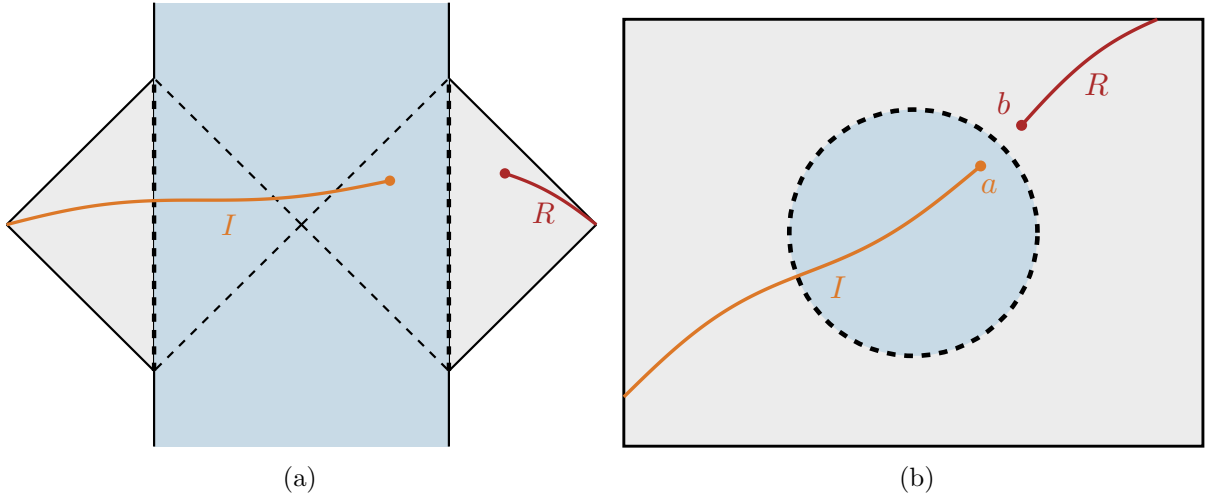
\begin{figure}[t]
\centering

\begin{subfigure}[t]{0.49\textwidth}
\centering
\resizebox{\linewidth}{!}{%
\begin{tikzpicture}[scale=2.6]

\definecolor{celestesave}{RGB}{200,218,230}
\definecolor{puntito}{RGB}{220,120,40}
\definecolor{rojito}{RGB}{170,40,40}

\def\ptDia{3.8pt}
\def\xmax{1.45}
\def\ymin{-0.4}
\def\ymax{1.8}

\pgfmathsetmacro{\yc}{(\ymin+\ymax)/2}
\pgfmathsetmacro{\xmid}{\xmax/2}
\pgfmathsetmacro{\Delta}{\xmax-\xmid}

\fill[celestesave] (0,\ymin) rectangle (\xmax,\ymax);

\fill[gray!15]
  ({\xmax+\Delta},\yc)
  -- (\xmax,{\yc+\Delta})
  -- (\xmax,{\yc-\Delta})
  -- cycle;

\draw[thick] ({\xmax+\Delta},\yc) -- (\xmax,{\yc+\Delta});
\draw[thick] ({\xmax+\Delta},\yc) -- (\xmax,{\yc-\Delta});
\draw[thick] (\xmax,{\yc+\Delta}) -- (\xmax,{\yc-\Delta});

\fill[gray!15]
  ({-\xmid},\yc)
  -- (0,{\yc+\Delta})
  -- (0,{\yc-\Delta})
  -- cycle;

\draw[thick] ({-\xmid},\yc) -- (0,{\yc+\Delta});
\draw[thick] ({-\xmid},\yc) -- (0,{\yc-\Delta});
\draw[thick] (0,{\yc+\Delta}) -- (0,{\yc-\Delta});

\fill[celestesave]
  (\xmid,\yc)
  -- (\xmax,{\yc+\Delta})
  -- (\xmax,{\yc-\Delta})
  -- cycle;

\draw[thick,dashed] (\xmid,\yc) -- (\xmax,{\yc+\Delta});
\draw[thick,dashed] (\xmid,\yc) -- (\xmax,{\yc-\Delta});

\fill[celestesave]
  (\xmid,\yc)
  -- (0,{\yc+\Delta})
  -- (0,{\yc-\Delta})
  -- cycle;

\draw[thick,dashed] (\xmid,\yc) -- (0,{\yc+\Delta});
\draw[thick,dashed] (\xmid,\yc) -- (0,{\yc-\Delta});

\draw[thick] (0,\ymin) -- (0,{\yc-\Delta});
\draw[thick] (0,{\yc+\Delta}) -- (0,\ymax);
\draw[thick] (\xmax,\ymin) -- (\xmax,{\yc-\Delta});
\draw[thick] (\xmax,{\yc+\Delta}) -- (\xmax,\ymax);

\draw[line width=1.6pt,dashed] (0,{\yc-\Delta}) -- (0,{\yc+\Delta});
\draw[line width=1.6pt,dashed] (\xmax,{\yc-\Delta}) -- (\xmax,{\yc+\Delta});

\pgfmathsetmacro{\xL}{-\xmid}
\pgfmathsetmacro{\yL}{\yc}

\pgfmathsetmacro{\xE}{\xmax+\Delta}
\pgfmathsetmacro{\yE}{\yc}

\pgfmathsetmacro{\xP}{\xmid + 0.6*(\xmax-\xmid)}
\pgfmathsetmacro{\yP}{\yc + 0.3*\Delta}

\pgfmathsetmacro{\xR}{\xmax + 0.38*\Delta}
\pgfmathsetmacro{\yR}{\yc + 0.35*\Delta}

\pgfmathsetmacro{\Aor}{0.035}
\draw[puntito, line width=1.2pt]
  plot[domain=0:1, samples=220, variable=\s]
  ({\xL + (\xP-\xL)*\s},
   {\yL + (\yP-\yL)*\s + 0.05*\s*(1-\s) + \Aor*sin(360*\s)});

\pgfmathsetmacro{\Are}{0.02}
\draw[rojito, line width=1.2pt]
  plot[domain=0:1, samples=220, variable=\s]
  ({\xE + (\xR-\xE)*\s},
   {\yE + (\yR-\yE)*\s + 0.04*\s*(1-\s) + \Are*sin(180*\s)});

\node[circle, fill=puntito, inner sep=0pt, minimum size=\ptDia, transform shape=false] at (\xP,\yP) {};
\node[circle, fill=rojito,  inner sep=0pt, minimum size=\ptDia, transform shape=false] at (\xR,\yR) {};

\pgfmathsetmacro{\sI}{0.58}
\pgfmathsetmacro{\xI}{\xL + (\xP-\xL)*\sI}
\pgfmathsetmacro{\yI}{\yL + (\yP-\yL)*\sI + 0.05*\sI*(1-\sI) + \Aor*sin(360*\sI)}
\node[puntito, font=\bfseries] at (\xI,{\yI-0.12}) {$I$};

\pgfmathsetmacro{\sRR}{0.58}
\pgfmathsetmacro{\xRR}{\xE + (\xR-\xE)*\sRR}
\pgfmathsetmacro{\yRR}{\yE + (\yR-\yE)*\sRR + 0.04*\sRR*(1-\sRR) + \Are*sin(180*\sRR)}
\node[rojito, font=\bfseries] at (\xRR,{\yRR-0.12}) {$R$};

\end{tikzpicture}%
}
\caption{}
\end{subfigure}%
\hfill
\begin{subfigure}[t]{0.49\textwidth}
\centering
\resizebox{\linewidth}{!}{%
\begin{tikzpicture}

\definecolor{celestesave}{RGB}{200,218,230}
\definecolor{puntito}{RGB}{220,120,40}
\definecolor{rojito}{RGB}{170,40,40}

\def\ptSize{1.9pt}
\def\R{1.5}
\def\W{7.0}
\def\H{5.2}

\fill[gray!15] (-\W/2,-\H/2) rectangle (\W/2,\H/2);
\draw[line width=1pt] (-\W/2,-\H/2) rectangle (\W/2,\H/2);

\filldraw[
    fill=celestesave,
    draw=black,
    line width=1.6pt,
    dashed
]
(0,0) circle (\R);

\pgfmathsetmacro{\xS}{-\W/2}
\pgfmathsetmacro{\yS}{-\H/2 + 0.6}
\pgfmathsetmacro{\inset}{0.35}

\pgfmathsetmacro{\ReffIn}{\R - \inset}
\pgfmathsetmacro{\xE}{\ReffIn*cos(45)}
\pgfmathsetmacro{\yE}{\ReffIn*sin(45)}

\pgfmathsetmacro{\ReffOut}{\R + \inset}
\pgfmathsetmacro{\xRed}{\ReffOut*cos(45)}
\pgfmathsetmacro{\yRed}{\ReffOut*sin(45)}

\fill[puntito] (\xE,\yE) circle (\ptSize);
\fill[rojito]  (\xRed,\yRed) circle (\ptSize);

\pgfmathsetmacro{\Aor}{0.18}
\draw[puntito, line width=1.2pt]
  plot[domain=0:1, samples=220, variable=\s]
  ({\xS + (\xE-\xS)*\s},
   {\yS + (\yE-\yS)*\s + 0.35*\s*(1-\s) + \Aor*sin(360*\s)});

\pgfmathsetmacro{\xTop}{\W/2 - 0.55}
\pgfmathsetmacro{\yTop}{\H/2}

\pgfmathsetmacro{\Are}{0.12}
\draw[rojito, line width=1.2pt]
  plot[domain=0:1, samples=220, variable=\s]
  ({\xRed + (\xTop-\xRed)*\s},
   {\yRed + (\yTop-\yRed)*\s + 0.20*\s*(1-\s) + \Are*sin(180*\s)});

\pgfmathsetmacro{\sI}{0.60}
\pgfmathsetmacro{\xI}{\xS + (\xE-\xS)*\sI}
\pgfmathsetmacro{\yI}{\yS + (\yE-\yS)*\sI + 0.35*\sI*(1-\sI) + \Aor*sin(360*\sI)}
\node[puntito, font=\bfseries] at (\xI,{\yI-0.33}) {$I$};

\pgfmathsetmacro{\sR}{0.55}
\pgfmathsetmacro{\xRR}{\xRed + (\xTop-\xRed)*\sR}
\pgfmathsetmacro{\yRR}{\yRed + (\yTop-\yRed)*\sR + 0.20*\sR*(1-\sR) + \Are*sin(180*\sR)}
\node[rojito, font=\bfseries] at (\xRR,{\yRR-0.38}) {$R$};

\node[puntito, font=\bfseries]
at ([xshift=3pt,yshift=-7pt]\xE,\yE) {$a$};

\node[rojito, font=\bfseries, above left]
at (\xRed,\yRed) {$b$};

\end{tikzpicture}
}
\caption{}
\end{subfigure}

\caption{Representative one-QES building block. (a) Lorentzian picture with a single radiation
interval $R$ and its associated island interval $I$. (b) Euclidean disk description of the same local
endpoint data. The finite-$n$ QES position is denoted by $a$, while $b$ denotes the bath endpoint
that enters the one-sided refined entropy formula.}
\label{fig:oneqes-setup}
\end{figure}

For the one-QES problem the relevant uniformization is
\begin{equation}
W(w)=\left(\frac{w-a}{1-\bar a\, w}\right)^{1/n},
\label{eq:oneqes-uniformization}
\end{equation}
where $a$ is the finite-$n$ QES position in the unit disk, determined dynamically by extremizing the generalized modular entropy. This is the finite-$n$ analogue of the standard one-QES geometry, and it is the local building block that controls the factorized late-time branch. The corresponding Lorentzian and Euclidean pictures are shown in
Fig.~\ref{fig:oneqes-setup}.

It is important to distinguish the replica uniformization map $W(w)$ from the welding maps
$F$ and $G$ that we will introduce in Section \ref{sec:welding}. The map $W$ uniformizes the local one-QES replica
geometry with a branch point at $a$, whereas $F$ and $G$ reconstruct the corresponding
physical replicated domain from the exterior and interior of the unit circle, respectively.

It is convenient to use the small parameter $\kappa$, introduced in
\eqref{eq:kappa-regime}, as the perturbative bookkeeping parameter. Using
\begin{equation}
\frac{\phi_r}{4\beta G_N}=\frac{c}{24\pi\kappa},
\label{eq:phir-kappa-relation}
\end{equation}
one sees that the boundary reparametrization is perturbatively close to the thermal circle when
$\kappa\ll1$. In practice this means that one may write
\begin{equation}
\theta(\tau)=\tau+\delta\theta(\tau),\qquad \delta\theta=O(\kappa),
\end{equation}
and solve the boundary equation iteratively in powers of $\kappa$. This is precisely the regime in
which the associated conformal welding problem becomes tractable.

\subsection{Perturbative expansion and the linearized \texorpdfstring{$\delta\theta$}{delta theta} equation}

We parameterize the boundary reparametrization as
\begin{equation}
\delta\theta(\tau)=\sum_{m\in\mathbb Z\setminus\{0,\pm1\}} c_m e^{im\tau}.
\label{eq:deltatheta-Fourier}
\end{equation}
The modes \(m=0,\pm1\) are absent because they are pure global
\(\mathrm{PSL}(2,\mathbb R)\) transformations. We further expand
\begin{equation}
\delta\theta(\tau)=\kappa\,\theta_1(\tau)+\kappa^2\,\theta_2(\tau)+O(\kappa^3).
\label{eq:deltatheta-kappa-expansion}
\end{equation}

To obtain the linearized equation, one first expands the Schwarzian piece around the trivial
boundary \(w=e^{i\tau}\). A direct computation gives
\begin{equation}
\{w,\tau\}
=
\frac12+\delta\theta'+\delta\theta'''
+\frac12(\delta\theta')^2-\delta\theta'\delta\theta'''
-\frac32(\delta\theta'')^2+O(\delta\theta^3).
\label{eq:Schwarzian-expanded}
\end{equation}
For the linear kernel it is enough to set \(a=0\) in \eqref{eq:oneqes-uniformization}, so that
\(W(w)=w^{1/n}\). This yields
\begin{equation}
T_0(w)=\frac{\varepsilon_n}{w^2},
\qquad
\varepsilon_n\equiv \frac{n^2-1}{4n^2}.
\label{eq:T0-def}
\end{equation}
Using \(w'=i(1+\delta\theta')w\), one finds that the linear part of \(B(\tau)\) is
\begin{equation}
B^{(1)}[\delta\theta]
=
\delta\theta'''+\frac{1}{n^2}\,\delta\theta'.
\label{eq:B-linear}
\end{equation}
Therefore the linearized boundary equation takes the form
\begin{equation}
\left(\partial_\tau^4+\frac{1}{n^2}\partial_\tau ^2\right)\delta\theta(\tau)
=
\frac{24\pi\kappa}{c}\,
i\Big(T_{yy}(i\tau)-T_{\bar y\bar y}(-i\tau)\Big)_{\rm lin}.
\label{eq:linear-deltatheta-eom}
\end{equation}
This is the basic equation for the replicated boundary mode. The right-hand side depends on
the one-QES replica data through the matter stress tensor, while the left-hand side is universal.

At the next order, the quadratic part of the Schwarzian expansion induces nonlinear mode mixing.
Although we will not need the full second-order equation explicitly, it is useful to note that the
universal quadratic contribution is
\begin{equation}
B^{(2)}[\delta\theta]
=
\frac{1}{2n^2}(\delta\theta')^2
-\delta\theta'\delta\theta'''
-\frac32(\delta\theta'')^2.
\label{eq:B-quadratic}
\end{equation}
This is the origin of the nonlinear source terms that first enter at \(O(\kappa^2)\).

For the one-QES saddle, the linearized equation \eqref{eq:linear-deltatheta-eom} can be solved
mode by mode. The result obtained in \cite{Almheiri2020,HKP2024} is
\begin{equation}
c_m=-i\,\kappa\,u_m(n)\,b^{-m},
\qquad m>1,
\label{eq:cm-general}
\end{equation}
with
\begin{equation}
u_m(n):=
\frac{n^2-1}{2}\,
\frac{m+1}{m^2\big((mn)^2-1\big)}.
\label{eq:um-def}
\end{equation}
Reality of the boundary curve fixes the negative-frequency coefficients accordingly.

Several comments are in order. First, the dependence \(c_m\sim b^{-m}\) exhibits the
late-time hierarchy directly: higher modes are increasingly suppressed at large \(|b|\).
Second, because \(m=0,\pm1\) are gauge, the first physical mode is \(m=2\). This is why the
dominant-mode truncation keeps precisely that mode. Third, this hierarchy is
the real justification for the truncation used later in the paper: one first works in the
factorized large-\(|b|\) regime, and only afterwards evaluates the resulting one-sided building
block on the representative \(|b|=1\).

For the leading mode one finds
\begin{equation}
u_2(n)=\frac{3(n^2-1)}{8(4n^2-1)}
\equiv \gamma(n),
\label{eq:gamma}
\end{equation}
so that
\begin{equation}
c_2=-i\,\gamma(n)\,\frac{\kappa}{b^2}.
\label{eq:c2-leading}
\end{equation}
This single coefficient controls all the \(O(\kappa)\) data needed in the main text.

The first omitted mode is \(m=3\), with coefficient
\begin{equation}
u_3(n)=\frac{2(n^2-1)}{9(9n^2-1)}
\equiv \eta(n),
\label{eq:eta}
\end{equation}
and is further suppressed by an additional factor of \(|b|^{-1}\) at the level of the boundary
deformation. More importantly, when translated into the one-sided modular-entropy building block,
its contribution is suppressed by an additional factor of \(|b|^{-2}\) relative to the leading
\(m=2\) term. We show this explicitly in Appendix~\ref{app:m3mode}. Thus the coefficient extracted
in the main text should be understood as the leading coefficient in the factorized large-\(|b|\)
hierarchy.

\subsection{Welding maps and one-QES refined entropy}{\label{sec:welding2}}

Once the boundary mode is known, the next step is to reconstruct the replicated geometry from the
deformed physical boundary. This is the conformal welding problem. One starts from the unit circle,
parametrized by $u=e^{i\tau}$, and from the corresponding deformed boundary
\begin{equation}
w_n(\tau)=e^{i\theta(\tau;n)} .
\end{equation}
The problem is to find two conformal maps: an interior map $G$, defined on the unit disk, and an
exterior map $F$, defined on its complement, such that both describe the same physical boundary.
Equivalently, their boundary values must satisfy the matching condition
\begin{equation}
G_n\!\left(w_n(\tau)\right)=F_n(u), \qquad |u|=1 .
\end{equation}
At linear order this becomes the additive boundary-value problem discussed in Appendix~\ref{app:nonfactorising}; in that
sense, conformal welding is the mechanism that turns the boundary reparametrization $\theta(\tau;n)$
into the uniformization data of the finite-$n$ saddle.

\begin{figure}[t]
\centering

\begin{tikzpicture}

\definecolor{celestesave}{RGB}{200,218,230}
\definecolor{elegantblue}{RGB}{35,70,120}
\definecolor{elegantgray}{RGB}{110,110,110}

\def\R{1.5cm}
\def\W{7.0cm}
\def\H{5.2cm}

\def\Xsep{8.8}
\pgfmathsetmacro{\Xmid}{\Xsep/2}

\pgfmathsetlengthmacro{\dxblue}{0.75*\R*cos(55)}
\pgfmathsetlengthmacro{\dxgray}{1.28*\R*cos(55)}
\pgfmathsetlengthmacro{\ygrayend}{1.35*\R*sin(-125)}

\def\Rin{0.85*\R}
\def\Rout{1.15*\R}

\def\angstartinner{-55}
\def\angendinner{15}
\def\angstartouter{-45}
\def\angendouter{5}

\def\anglabelblue{-10}
\def\Rtextblue{0.52*\R}
\def\anglabelgray{-20}
\def\Rtextgray{1.42*\R}

\begin{scope}

  \fill[gray!15] (-\W/2,-\H/2) rectangle (\W/2,\H/2);
  \draw[line width=1pt] (-\W/2,-\H/2) rectangle (\W/2,\H/2);

  \filldraw[
      fill=celestesave,
      draw=black,
      line width=1.6pt,
      dashed
  ]
  (0,0) circle (\R);

  \draw[elegantblue,line width=2pt,-{Stealth[length=3.2mm,width=2.4mm]}]
    ([shift={(\angstartinner:\Rin)}]0,0)
    arc[start angle=\angstartinner, end angle=\angendinner, radius=\Rin];

  \draw[elegantgray,line width=1.8pt,-{Stealth[length=3.2mm,width=2.4mm]}]
    ([shift={(\angstartouter:\Rout)}]0,0)
    arc[start angle=\angstartouter, end angle=\angendouter, radius=\Rout];

  \node[elegantblue]
    at ([shift={(\anglabelblue:\Rtextblue-0.5)}]0,0)
    {$\mathbf{\boldsymbol{e^{i\theta(\tau)}}}$};

  \node[elegantgray]
    at ([shift={(\anglabelgray:\Rtextgray)}]0,0)
    {$\mathbf{\boldsymbol{e^{i\tau}}}$};

  \node at ([shift={(135:0.48*\R)}]0,0)
    {$|w|<1$};

  \node[anchor=south west]
    at (-\W/2+0.2,-\H/2+0.2)
    {$|v|>1$};

\end{scope}

\begin{scope}[xshift=\Xsep cm]

  \fill[gray!15] (-\W/2,-\H/2) rectangle (\W/2,\H/2);
  \draw[line width=1pt] (-\W/2,-\H/2) rectangle (\W/2,\H/2);

  \filldraw[
      fill=celestesave,
      draw=black,
      line width=1.6pt,
      dashed,
      smooth
  ]
  plot[samples=360, domain=0:360]
  (
    {(\R*(1 + 0.18*(0.65*sin(3*\x) + 0.45*sin(7*\x+25) + 0.35*sin(11*\x+80))))*cos(\x)},
    {(\R*(1 + 0.18*(0.65*sin(3*\x) + 0.45*sin(7*\x+25) + 0.35*sin(11*\x+80))))*sin(\x)}
  )
  -- cycle;

  \node[anchor=south east]
    at (\W/2-0.2,-\H/2+0.2)
    {\large $z$};

\end{scope}

\draw[
  elegantblue,
  line width=2.6pt,
  -{Stealth[length=5mm,width=3.5mm]}
]
  ([shift={(55:0.75*\R)}]0,0)
  .. controls (2.8,2.8) and (5.0,2.3)
  .. ([shift={(-\dxblue,0.6cm)}]\Xsep cm,0);

\node[elegantblue] at (\Xmid cm,1.35cm) {\large$\displaystyle G(w)$};

\draw[
  black!70,
  line width=2.6pt,
  -{Stealth[length=5mm,width=3.5mm]}
]
  ([shift={(-55:1.28*\R)}]0,0)
  .. controls (2.8,-2.8) and (5.4,-2.3)
  .. ([shift={(-\dxgray,\ygrayend)}]\Xsep cm,0);

\node[black!70] at (\Xmid cm,-1.55cm) {\large$\displaystyle F(v)$};

\end{tikzpicture}

\caption{Conformal welding for the one-QES replicated geometry. The interior map $G$ and the
exterior map $F$ uniformize the same deformed physical boundary from the inside and outside,
respectively.}
\label{fig:welding-conforme}

\end{figure}
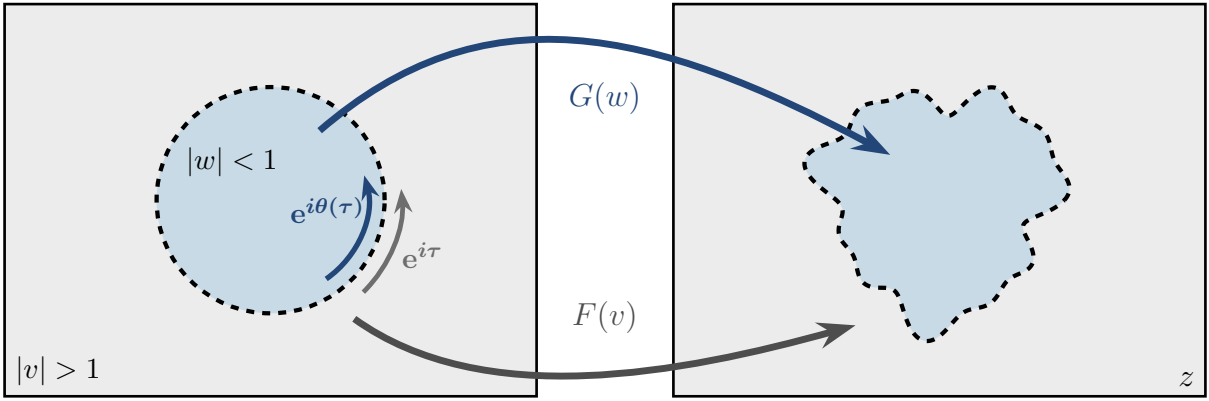

Figure~\ref{fig:welding-conforme} illustrates this structure. The map $G$ sends the unit disk to the
physical replicated region from the inside, while $F$ does the same from the outside. Once these maps
are known, the matter refined entropy of the one-QES saddle can be written directly in terms of the
bath endpoint $b$, the finite-$n$ QES position $a$, and the corresponding uniformization data:
\begin{equation}
\tilde S^{\rm CFT}_{n}(a,\bar a)=
\frac{c}{6n}
\log
\frac{|F(b)-G(a)|^2}
{(1-|a|^2)\,|G'(a)F'(b)b|}.
\label{eq:oneqes-refined-entropy}
\end{equation}
Here $a$ and $b$ are the local endpoint data shown in Fig.~\ref{fig:oneqes-setup}(b): $a$ is the finite-$n$ QES position in the unit disk, while $b$ denotes the image of the bath endpoint in the exterior $z$-plane. In the factorized treatment used throughout this paper, the parameter $b$ is inherited from the embedding of the one-sided building block into the full two-sided geometry, and the large-$|b|$ hierarchy is the bookkeeping device controlling that embedding.

The corresponding generalized refined entropy is
\begin{equation}
\Stgen_n(a,\bar a)
=
\frac{\Phi^{(n)}(a,\bar a)}{4G_N}
+
\St_n^{\CFT}(a,\bar a),
\label{eq:oneQES-generalized}
\end{equation}
to be extremized with respect to \(a\) at fixed \(n\). This is the one-sided finite-$n$
functional whose doubled version governs the factorized two-sided island branch.

At linear order in the boundary deformation, the welding maps admit expansions of the form
\begin{equation}
G(w)=w+O(c_m\,w^{m+1}),
\qquad
F(v)=v+O(c_{-m}\,v^{1-m}),
\label{eq:welding-general-structure}
\end{equation}
and in the dominant \(m=2\) truncation they simplify to
\begin{align}
G(w)
&=
w-\gamma(n)\,\frac{\kappa}{b^2}\,w^3
+O\!\left(\kappa|b|^{-3}w^4,\kappa^2\right),
\label{eq:G-leading}
\\
F(v)
&=
v-\gamma(n)\,\frac{\kappa}{\bar b^2}\,v^{-1}
+O\!\left(\kappa|b|^{-3}v^{-2},\kappa^2\right).
\label{eq:F-leading}
\end{align}
The precise form of the first omitted term is discussed in Appendix~\ref{app:m3mode}; for the
main computation it is enough to know that it is parametrically subleading in the factorized
expansion.

We can now summarize the one-sided data that will actually enter in Section~\ref{sec:main}. In the
factorized late-time regime, the full two-sided island branch reduces to
\begin{equation}
\Stgen_n(I)
=
\frac{4S_0}{n+1}
+
2\,\widetilde S^{\mathrm{gen}, \rm 1QES}_n\big|_{|b|=1}
+
O(e^{-2t_0}),
\label{eq:factorized-branch-reduction}
\end{equation}
where the first omitted terms are the genuinely non-factorizing inter-QES corrections. Their
expected parametric size is discussed in Appendix~\ref{app:nonfactorising}.

Before setting \(|b|=1\), the dominant one-sided quantities scale as
\begin{align}
F(b)
&=
b\Bigl(1-\gamma(n)\,\kappa |b|^{-4}
+O(\kappa |b|^{-6},\kappa^2)\Bigr),
\label{eq:Fb-general-scaling}
\\
F'(b)
&=
1+\gamma(n)\,\kappa |b|^{-4}
+O(\kappa |b|^{-6},\kappa^2),
\label{eq:Fpb-general-scaling}
\\
a
&=
\frac{\kappa}{\bar b}
+O(\kappa^2),
\label{eq:a-general-scaling}
\\
G(a)
&=
a+O(\kappa^3),
\qquad
G'(a)=1+O(\kappa^3).
\label{eq:Ga-general-scaling}
\end{align}
The last line follows because \(a=O(\kappa)\), while the first nonlinear correction to \(G\) in
the dominant truncation starts at cubic order in the boundary variable. Here $F'(b)$ ($G'(a)$) denotes the holomorphic derivative of the exterior welding map with respect to its argument $v$($w$), evaluated at $v=b$ ($w=a$) respectively.

These formulas already show why the main computation is simple. To obtain the leading
\(O(\kappa^2)\) factorized correction, one only needs the \(O(\kappa)\) pieces of \(F(b)\) and
\(F'(b)\), together with the first two terms in the QES position \(a\) once \(|b|=1\) is chosen.
Everything else is either higher order in \(\kappa\) or additionally suppressed in the large-\(|b|\) hierarchy. Since the genuinely inter-QES effects are parametrically suppressed in the late-time factorized regime, the leading contribution is captured by the local one-sided welding data around each QES, rather than by the full two-QES solution. Section~\ref{sec:main} is therefore reduced to a controlled expansion of using precisely this truncated one-sided data.

\section{The $O(\kappa^2)$ island coefficient}
\label{sec:main}

We now turn to the main question of the paper. In the late-time factorized regime the two-sided island saddle reduces, to the order retained here,
to two identical one-sided building blocks. The natural question is then whether this local
one-QES block already carries finite-$n$ information that is invisible in the $n\to1$ entropy
limit, or whether such information only appears once genuinely two-sided, non-factorizing
effects are included.

The result of this section is that the factorized saddle already does contain such information.
More precisely, the one-sided block develops a nontrivial $O(\kappa^2)$ finite-$n$ coefficient
whose value vanishes at $n=1$ but whose replica derivative does not. As a consequence, the
von Neumann entropy plateau is unchanged at this order, while the capacity of entanglement
receives a definite shift. In this sense, the capacity probes a nearby replica structure of the
island branch that the entropy itself does not detect.

Technically, the computation is simple once the perturbative ingredients of Section \ref{sec:welding} are in
place. The point, however, is not merely computational. What we are extracting below is the
first local/factorized imprint of replica backreaction on the island branch beyond the entropy
limit.

\subsection{Factorization and reduction to the representative one-sided block}

In the regime
\begin{equation}
e^{-t_0}\ll \kappa \ll 1,
\end{equation}
the left and right replicated geometries decouple to the order retained here, and the
two-sided island branch is obtained by doubling a one-QES building block and restoring the
topological contribution. Accordingly, the Hawking and island branches take the schematic form
\begin{align}
\tilde S^{\rm gen}_n(H)&=\frac{c}{3n}\left(
t_0-2\kappa\,\gamma(n)
+O(\kappa^2 |b|^{-8},\,\kappa^3,\,e^{-2t_0})
\right),\label{Shawking}\\[4pt]
\tilde S^{\rm gen}_n(I) &=
\frac{4S_0}{n+1}
+\frac{c}{6n}\left(
\frac{1}{\kappa}-2\kappa
-6\kappa\,\gamma(n)
+\alpha_I(n)\kappa^2
+O(\kappa^3)
\right).\label{islandS}
\end{align}
A direct expansion of the Hawking branch within the same dominant-mode truncation does
produce a quadratic term, but in the original large-$|b|$ hierarchy it appears only as
\begin{equation}
\delta \tilde S^{\rm gen}_n(H)\Big|_{\kappa^2}
=
\frac{c}{3n}\,\frac{\gamma(n)^2}{4}\,\kappa^2 |b|^{-8}
+ \cdots .
\end{equation}
This $O(\kappa^2 |b|^{-8})$ term has the same perturbative origin as the $\gamma(n)^2$ contribution that appears below in the island coefficient. The difference is that on the Hawking branch it remains subleading in the original large-$|b|$ factorized hierarchy and, moreover, since $\gamma(1)=0$, one has $\gamma(n)^2=O((n-1)^2)$ near $n=1$. It therefore does not contribute at order $\kappa^2$ either to the von Neumann entropy or to the capacity.

The entropy only probes the value of the branch at $n=1$, whereas $\alpha_I(n)$ encodes how
the finite-$n$ island saddle begins to depart from that limit. This is precisely the kind of data
to which the capacity is sensitive.
To extract $\alpha_I(n)$ it is enough to evaluate the representative one-sided block at $|b|=1$. In the explicit computation below we choose the symmetric representative $b=\bar b=1$,
for which the finite-$n$ QES position is real to the required order.
With this, the dominant-mode truncation reviewed in Section \ref{sec:welding} gives
\begin{align}
F(b) &= 1-\gamma(n)\kappa + O(\kappa^2),\\
F'(b) &= 1+\gamma(n)\kappa + O(\kappa^2),\\
a &= \kappa - 5\gamma(n)\kappa^2 + O(\kappa^3),\\
|a|^2 &= \kappa^2 + O(\kappa^3),
\end{align}
with \(\gamma(n)\) as defined in \eqref{eq:gamma}.
For the interior map, the only facts needed at this order are
\begin{equation}
G(a)=a+O(\kappa^3),\qquad G'(a)=1+O(\kappa^3).
\end{equation}
As discussed in Appendix \ref{app:m3mode}, the first omitted mode $m=3$ only contributes at subleading order
in the same factorized hierarchy and does not affect the leading coefficient extracted below.

\subsection{From the one-sided block to the factorized island branch}

We begin with the exact one-QES matter formula
\begin{equation}
\tilde S^{\rm CFT}_n(a,\bar a)=\frac{c}{6n}\log\frac{|F(b)-G(a)|^2}{(1-|a|^2)|G'(a)F'(b)b|}.
\end{equation}
For the representative symmetric block we choose $b=\bar b=1$
for which the finite-$n$ QES position is real to the order relevant here. We may therefore drop the bars and absolute values when expanding through $O(\kappa^2)$.
Inserting the truncated one-sided data gives
\begin{equation}
\frac{6n}{c}\tilde S^{\rm CFT}_n
=
\log\frac{(F(b)-a)^2}{(1-a^2)F'(b)}
+O(\kappa^3),
\end{equation}
and hence
\begin{equation}
\frac{6n}{c}\tilde S^{\rm CFT}_n
=
-(2+3\gamma (n))\kappa
+
\alpha_{\rm mat}(n)\kappa^2
+O(\kappa^3),
\end{equation}
with
\begin{equation}
\alpha_{\rm mat}(n)=8\gamma(n)-\frac12\gamma(n)^2.
\end{equation}
To obtain the coefficient of the generalized modular entropy, one must add the on-shell dilaton contribution. Evaluating the one-QES dilaton formula of \cite{HKP2024} on the representative symmetric block $b=\bar b=1$, and using the finite-$n$ QES position given above, one finds
\begin{equation}
\frac{6n}{c}\frac{\Phi^{(n)}(a)}{4G_N}
=
\frac{1}{2\kappa}
+\kappa
-4\gamma(n)\kappa^2
+O(\kappa^3).
\end{equation}
Therefore the one-sided generalized modular entropy becomes 
\begin{equation}
\frac{6n}{c}\tilde S^{\rm gen,1QES}_n
=
\frac{1}{2\kappa}
-(1+3\gamma (n))\kappa
+\frac{\alpha_I(n)}{2}\kappa^2
+O(\kappa^3),
\end{equation}
with
\begin{equation}
\begin{aligned}
\alpha_I(n)&=2(\alpha_{\rm mat}(n)-4\gamma(n))
=8\gamma(n)-\gamma(n)^2
\end{aligned}
\end{equation}
This is the main analytic result. Its significance lies not only in the existence of a nonzero
$O(\kappa^2)$ coefficient, but in its very specific replica behaviour:
\begin{equation}
\alpha_I(1)=0,\qquad \alpha_I'(1)=2.
\end{equation}
Thus the first local finite-$n$ backreaction of the factorized island branch is arranged in such a
way that it disappears exactly in the entropy limit, while remaining visible to observables that
differentiate with respect to $n$. This is the basic mechanism behind the capacity shift derived
below. Appendix~\ref{app:numerics-alpha} provides an independent numerical extraction of $\alpha_{mat}(n)$, and Appendix~\ref{app:numerics-derivative} checks the replica derivative at $n=1$.

Returning now to the full eternal black hole, doubling the one-sided contribution and reinstating the
topological term reproduces the factorized island branch written in \eqref{islandS}. The terms up to
$O(\kappa)$ agree with the factorized result \cite{HKP2024}, while the genuinely new information is the
explicit finite-$n$ coefficient $\alpha_I(n)$ multiplying $\kappa^2$.

It is worth emphasizing how this result should be interpreted. Equation \eqref{islandS} does not yet
solve the full two-sided replica problem beyond factorization. Rather, it isolates the leading
finite-$n$ correction intrinsic to the local one-QES building block from which the factorized
island branch is assembled. In that sense, it identifies the first nearby replica datum that is
already present before any genuinely non-factorizing inter-QES effects are included.

The capacity along a fixed saddle branch is
\begin{equation}
C^{(n)}=-n\partial_n \tilde S^{\rm gen}_n .
\end{equation}
As explained in Appendix~\ref{app:equivalence}, differentiating the on-shell branch is equivalent to the standard
dilaton-plus-matter formula once the finite-$n$ extremality condition is imposed.

Applying this to the island branch and evaluating at $n=1$, we obtain 
\begin{equation}
C_I=
S_0
+
\frac{c}{6\kappa}
-\frac{c}{12}\kappa
-\frac{c}{6}\alpha_I'(1)\kappa^2
+O(\kappa^3),
\end{equation}
and therefore
\begin{equation}
C_I=
S_0+\frac{c}{6\kappa}
-\frac{c}{12}\kappa
-\frac{c}{3}\kappa^2
+O(\kappa^3).
\end{equation}
Thus, relative to the result truncated through $O(\kappa)$, the island-branch capacity is shifted by
\begin{equation}
-\frac{c}{3}\kappa^2+O(\kappa^3).
\end{equation}

The contrast with the entropy is now completely transparent. Since $\alpha_I(1)=0$, the von
Neumann entropy plateau is unchanged at this order within the factorized approximation. By
contrast, since $\alpha_I'(1)=2\neq 0$, the capacity does feel the first variation of the finite-$n$ island branch around $n=1$.

\section{Conclusions}
\label{sec:conclusions}

In this paper we studied the capacity of entanglement on the island branch of the eternal JT black hole coupled to nongravitating baths, in the same controlled high-temperature regime used by Hollowood, Kumar and Piper \cite{HKP2024}. Our goal was deliberately focused. Rather than attempting to solve the full two-sided finite-$n$ replica problem, we asked a narrower question: whether the \emph{factorized} island saddle already carries finite-$n$ information that is invisible to the von Neumann entropy but detectable by the capacity. The answer, within the regime analysed here, is affirmative.

The technical result is the explicit determination of the first nontrivial $O(\kappa^2)$ correction to
the factorized island branch of the generalized modular entropy. The corresponding coefficient can
be obtained in closed form, but its real significance is not merely that it is nonzero. Rather, it
vanishes at $n=1$ while having a nonzero replica derivative there. As a consequence, within the
factorized approximation the late-time von Neumann entropy plateau is unchanged at order
$O(\kappa^2)$, whereas the late-time capacity plateau is shifted by a definite amount,
$-(c/3)\kappa^2 + O(\kappa^3)$.

In other words, the first nearby finite-$n$ information of the factorized island branch is invisible
to the entropy, but not to the capacity.

This is the main physical message of the paper. In the semiclassical gravity literature, much of the conceptual power of the replica method has come from the \(n\to 1\) limit, where generalized entropy and quantum extremality determine the fine-grained entropy of Hawking radiation \cite{LewkowyczMaldacena2013,FaulknerLewkowyczMaldacena2013,EngelhardtWall2015,Dong2016}. However, finite-$n$ saddles generally contain more information than survives in that limit. The capacity of entanglement is one natural observable that can access this nearby replica structure \cite{NakaguchiNishioka2016,deBoerJarvelaKeskiVakkuri2019}. Our calculation provides a concrete example in JT gravity where this extra sensitivity can be exhibited analytically: the factorized island saddle already distinguishes entropy from capacity at order \(O(\kappa^2)\), even before one includes genuinely non-factorizing two-sided effects.

From this viewpoint, our result helps separate two questions that are often entangled in
discussions of finite-$n$ islands. The first is whether the local, factorized island building block
already knows about nontrivial finite-$n$ data. The second is whether the complete two-sided
answer requires additional global information that only appears once factorization breaks down.
The analysis of this paper answers the first question in the affirmative, while the second remains
open. In particular, we do not determine the first inter-QES corrections that mix the two
asymptotic regions. As discussed parametrically in Appendix~\ref{app:nonfactorising}, those corrections are expected
to be suppressed in the hierarchy $e^{-t_0}\ll \kappa \ll 1$, but a full treatment would require
solving the genuinely two-sided problem beyond the factorized approximation. What we have
obtained is therefore the leading $O(\kappa^2)$ correction intrinsic to the factorized island branch
itself. This is already nontrivial, because the same branch exhibits no correction at this order in
the entropy. Put differently, the factorized saddle is more informative than the entropy plateau
alone would suggest, and the capacity functions here as a probe of the local replica structure of
the island branch rather than merely as a derived quantity attached to the $n=1$ entropy.

It is also worth emphasizing that the present result is not an artefact of an inconsistent truncation. The dominant-mode approximation used here is the same one that underlies the controlled factorized expansion of \cite{HKP2024}. Moreover, Appendix~\ref{app:m3mode} shows explicitly that the first omitted welding mode, \(m=3\), contributes only at subleading order in the same large-\(|b|\) hierarchy. Thus the coefficient \(\alpha_I(n)\) extracted in Section~\ref{sec:main} should be understood as the leading coefficient of the factorized expansion, not merely as the outcome of evaluating a prematurely truncated expression at \(|b|=1\).

There are several natural directions for future work. The most immediate one is to go beyond the factorized approximation and determine the first non-factorizing correction to the two-sided replica geometry. This would clarify whether the complete late-time capacity contains genuinely global contributions that are absent in the entropy at the same order. A second direction is to understand whether the mechanism found here is special to the high-temperature JT setup or whether it is more universal across two-dimensional dilaton gravity models with islands. In that regard, it would be particularly interesting to compare more systematically with evaporating backgrounds and with models such as RST, where the replica saddle may be sensitive to a broader set of dynamical scales \cite{AriasFondevila2026,HollowoodKumar2020,GotoHartmanTajdini2021}. A third direction is to ask whether the qualitative separation found here persists for finite-$n$ observables beyond the capacity itself. In particular, it would be interesting to understand whether similar effects arise for more general quantities built from modular-Hamiltonian moments or polynomials, of which the capacity is only the simplest example \cite{Arias2023}.

More conceptually, our analysis suggests that the structure of island saddles should not be judged solely through the lens of the \(n=1\) entropy. The entropy remains the quantity most directly tied to the Page curve and to the fine-grained information problem, but finite-$n$ data contain additional information about how the replica saddle is assembled. In the present example, that information is already present in the factorized one-QES building block, and the capacity is the first observable that makes it visible. This is perhaps the clearest way to summarize what we have learned: even in a regime where the entropy plateau looks rigid, the nearby replica geometry is not.

\section*{Acknowledgments}
We would like to thank to Dani Fondevila and Prem Kumar for comments on a draft. AT is supported by fellowships from CONICET (Argentina).

\appendix

\section{Replica extremality and the capacity formula}
\label{app:equivalence}

In the main text we computed the capacity by differentiating the on-shell branch
\begin{equation}
C^{(n)}=-\,n\,\partial_n \Stgen_n,
\qquad
C=C^{(n)}\big|_{n=1}.
\label{eq:appA-capacity-branch}
\end{equation}
Since the island saddle is defined by a finite-$n$ extremization problem, it is natural to ask
why this prescription is equivalent to the more standard formula in which the capacity is written
as an explicit derivative of the dilaton term plus a matter contribution, both evaluated at the
finite-$n$ QES \cite{KawabataNishiokaOkuyamaWatanabe2021}. The answer is simple: once the generalized modular entropy is extremized with
respect to the QES position at fixed replica index, differentiating the on-shell branch is
equivalent to differentiating the off-shell functional at fixed QES position. In mathematical
terms, this is just the envelope theorem applied to the finite-$n$ extremization problem.

This statement is closely analogous to the usual logic behind gravitational replica calculations:
one first defines a finite-$n$ generalized entropy functional, then determines the replica surface
by extremization, and only afterwards studies its $n$-dependence
\cite{LewkowyczMaldacena2013,EngelhardtWall2015,Dong2016,HKP2024}.
For refined entropy and capacity in JT gravity, the same structure underlies the formulas used in
\cite{KawabataNishiokaOkuyamaWatanabe2021,HKP2024}. Since our main text uses the derivative of the
on-shell branch directly, it is useful to spell out this equivalence explicitly.

\subsection{Off-shell functional and finite-$n$ QES}
\label{appA:offshell}

Let \(\{w_i,\bar w_i\}\) denote the endpoints of the candidate island region in a given replica
geometry. The discussion below applies both to the one-QES building block and to more general
multi-QES configurations; in the former case the index \(i\) takes a single value.
At fixed replica index \(n\), define the off-shell generalized refined entropy functional
\begin{equation}
\Stgen(n;\{w_i,\bar w_i\})
=
S_{\rm top}(n)
+
\sum_{i\in\partial I}
\frac{\Phi^{(n)}(w_i,\bar w_i)}{4G_N}
+
\St_n^{\rm mat}(n;\{w_i,\bar w_i\}),
\label{eq:appA-offshel-functional}
\end{equation}
where \(S_{\rm top}(n)\) denotes any branch-dependent topological contribution independent of the
QES positions. For the factorized two-sided island branch studied in the main text,
\begin{equation}
S_{\rm top}(n)=\frac{4S_0}{n+1},
\label{eq:appA-stop}
\end{equation}
while for the one-sided building block or for the Hawking branch one may simply omit this term.

The finite-$n$ QES positions \(w_{\star,i}(n)\) are determined by extremizing
\eqref{eq:appA-offshel-functional} at fixed \(n\),
\begin{equation}
\partial_{w_i}\Stgen(n;\{w_j,\bar w_j\})\Big|_{w=w_\star(n)}=0,
\qquad
\partial_{\bar w_i}\Stgen(n;\{w_j,\bar w_j\})\Big|_{w=w_\star(n)}=0,
\qquad
\forall i.
\label{eq:appA-finite-n-extremality}
\end{equation}
The corresponding on-shell generalized refined entropy is then
\begin{equation}
\Stgen_n
=
\Stgen\bigl(n;\{w_{\star,i}(n),\bar w_{\star,i}(n)\}\bigr).
\label{eq:appA-onshell-def}
\end{equation}

Equation \eqref{eq:appA-finite-n-extremality} is the finite-$n$ analogue of the usual QES
condition. The only difference from the more familiar \(n=1\) case is that both the dilaton and
the matter refined entropy are evaluated on an \(n\)-dependent replica background, so the
extremal surface itself also depends on \(n\). This is precisely the structure emphasized in the
main text: the capacity probes the first nontrivial variation of this finite-$n$ saddle.

\subsection{Envelope theorem for the replica branch}
\label{appA:envelope}

We now differentiate the on-shell quantity \eqref{eq:appA-onshell-def} with respect to \(n\).
A direct application of the chain rule gives
\begin{align}
\partial_n \Stgen_n
&=
\partial_n
\Stgen\bigl(n;\{w_i,\bar w_i\}\bigr)\Big|_{w=w_\star(n)}
\nonumber\\[1mm]
&\quad
+
\sum_i
\partial_{w_i}\Stgen\bigl(n;\{w_j,\bar w_j\}\bigr)\Big|_{w=w_\star(n)}
\,\partial_n w_{\star,i}(n)
\nonumber\\[1mm]
&\quad
+
\sum_i
\partial_{\bar w_i}\Stgen\bigl(n;\{w_j,\bar w_j\}\bigr)\Big|_{w=w_\star(n)}
\,\partial_n \bar w_{\star,i}(n).
\label{eq:appA-chain-rule}
\end{align}
The last two lines vanish identically by the finite-$n$ extremality condition
\eqref{eq:appA-finite-n-extremality}. Therefore
\begin{equation}
\partial_n \Stgen_n
=
\partial_n
\Stgen\bigl(n;\{w_i,\bar w_i\}\bigr)\Big|_{w=w_\star(n)}.
\label{eq:appA-envelope}
\end{equation}
This is the precise statement of the envelope theorem in the present context.

The meaning of \eqref{eq:appA-envelope} is simple but important. Once the finite-$n$ QES has
been imposed, the derivative of the \emph{on-shell} branch is obtained by differentiating the
off-shell functional while keeping the QES position fixed. In particular, there is no additional
contribution proportional to \(\partial_n w_{\star,i}\). This is why one may safely compute the
capacity by differentiating the branch expression written in the main text.

It is worth stressing that this does \emph{not} mean one is extremizing the capacity itself.
The extremization problem is the one that defines the finite-$n$ generalized refined entropy.
The capacity is then obtained by differentiating the corresponding on-shell branch, exactly as in
\eqref{eq:appA-capacity-branch}.

\subsection{Capacity formula}
\label{appA:capacity}

Using \eqref{eq:appA-envelope} inside \eqref{eq:appA-capacity-branch}, one finds
\begin{equation}
C^{(n)}
=
-\,n\,\partial_n
\Stgen\bigl(n;\{w_i,\bar w_i\}\bigr)\Big|_{w=w_\star(n)}.
\label{eq:appA-Cn-fixedQES}
\end{equation}
Substituting the decomposition \eqref{eq:appA-offshel-functional}, this becomes
\begin{equation}
C^{(n)}
=
-\,n\,\partial_n S_{\rm top}(n)
-
n\sum_{i\in\partial I}
\frac{\partial_n \Phi^{(n)}(w_i,\bar w_i)}{4G_N}\Big|_{w=w_\star(n)}
-
n\,\partial_n \St_n^{\rm mat}(n;\{w_i,\bar w_i\})\Big|_{w=w_\star(n)}.
\label{eq:appA-Cn-decomposed}
\end{equation}
This is the general finite-$n$ form of the ``dilaton formula'' alluded to in the main text:
the capacity is the sum of a topological contribution, an explicit replica derivative of the
dilaton term, and an explicit replica derivative of the matter refined entropy, all evaluated at
the finite-$n$ extremal surface.

At \(n=1\), one obtains
\begin{equation}
C
=
C_{\rm top}
-
\sum_{i\in\partial I}
\frac{\partial_n \Phi^{(n)}(w_i,\bar w_i)}{4G_N}\Big|_{n=1,w=w_\star(1)}
+
C_{\rm mat},
\label{eq:appA-C1}
\end{equation}
with
\begin{equation}
C_{\rm top}
:=
-n\,\partial_n\!\left[S_{\rm top}(n)\right]\Big|_{n=1},
\qquad
C_{\rm mat}
:=
-\,\partial_n \St_n^{\rm mat}(n;\{w_i,\bar w_i\})\Big|_{n=1,w=w_\star(1)}.
\label{eq:appA-Ctop-Cmat}
\end{equation}
For the factorized two-sided island branch of the main text, where \(S_{\rm top}(n)=4S_0/(n+1)\),
this gives
\begin{equation}
C_{\rm top}=S_0.
\label{eq:appA-Ctop-island}
\end{equation}

The conclusion is therefore straightforward. The two procedures are equivalent:

\begin{enumerate}
\item first extremize \(\Stgen(n;\{w_i,\bar w_i\})\) at fixed \(n\), obtain the on-shell branch
\(\Stgen_n\), and then differentiate with respect to \(n\);

\item or first use the envelope theorem to differentiate the off-shell functional at fixed QES
position, which produces the explicit dilaton-plus-matter formula \eqref{eq:appA-Cn-decomposed}.
\end{enumerate}

Once the finite-$n$ extremality condition is imposed, both routes give exactly the same answer.
This is the sense in which the capacity formula used in the main text is completely consistent
with the standard QES logic of gravitational replica calculations.

\section{Linearized welding and parametric size of inter-QES corrections}\label{app:nonfactorising}

In the main text we worked in the factorized late-time regime
\begin{equation}
e^{-t_0}\ll \kappa \ll 1 ,
\label{B.1}
\end{equation}
and consistently discarded the first genuinely non-factorizing two-sided corrections. The
purpose of this appendix is not to solve the full two-QES problem, but to make more explicit
why, under a natural regularity assumption, the omitted inter-QES contribution is expected to
be parametrically smaller than the factorized result computed in Section~4. The discussion is
therefore complementary to the main text: it clarifies the structure of the linearized welding
problem near $n=1$, and explains why the first non-factorizing correction should enter only at
subleading order in the same late-time hierarchy.

\subsection{Linearized welding near \texorpdfstring{$n=1$}{n=1}}

Let
\begin{equation}
u \equiv e^{i\tau}, \qquad |u|=1 ,
\label{B.2}
\end{equation}
and consider the replicated physical boundary
\begin{equation}
w_n(\tau)=e^{i\theta(\tau;n)} .
\label{B.3}
\end{equation}
To isolate the first variation relevant for the capacity, we expand simultaneously in $\kappa$
and in $n-1$:
\begin{equation}
w_n(\tau)=u+\kappa (n-1)\,\xi(\tau)
+O\!\left(\kappa (n-1)^2,\kappa^2\right),
\qquad
\xi(\tau)=i\,u\,\eta_1(\tau),
\label{B.4}
\end{equation}
where $\eta_1(\tau)$ is a real $2\pi$-periodic function. Equivalently,
\begin{equation}
\theta(\tau;n)=\tau+\kappa (n-1)\,\eta_1(\tau)
+O\!\left(\kappa (n-1)^2,\kappa^2\right).
\label{B.5}
\end{equation}

We now expand the interior and exterior welding maps as
\begin{equation}
G_n(w)=w+\kappa (n-1)\,g(w)
+O\!\left(\kappa (n-1)^2,\kappa^2\right),
\qquad
F_n(u)=u+\kappa (n-1)\,f(u)
+O\!\left(\kappa (n-1)^2,\kappa^2\right).
\label{B.6}
\end{equation}
The exact matching condition on the unit circle,
\begin{equation}
G_n\!\left(w_n(\tau)\right)=F_n(u),
\label{B.7}
\end{equation}
then reduces at first non-trivial order to
\begin{equation}
g_+(u)-f_-(u)=-\xi(u), \qquad |u|=1,
\label{B.8}
\end{equation}
where $g_+$ and $f_-$ denote the boundary values taken from inside and outside the unit circle,
respectively. Thus the linearized welding problem is an additive Riemann--Hilbert problem on
the circle.

Introducing the Hardy projectors
\begin{equation}
P_+\phi(u)=\sum_{m\ge 0}\phi_m u^m,
\qquad
P_-\phi(u)=\sum_{m<0}\phi_m u^m,
\qquad
\phi(u)=\sum_{m\in\mathbb Z}\phi_m u^m,
\label{B.9}
\end{equation}
a convenient normalized solution of \eqref{B.8} is
\begin{equation}
g_+(u)=-P_+\xi(u),
\qquad
f_-(u)=P_-\xi(u).
\label{B.10}
\end{equation}
Indeed,
\begin{equation}
g_+(u)-f_-(u)
=
-\,P_+\xi(u)-P_-\xi(u)
=
-\xi(u).
\label{B.11}
\end{equation}
This is the linearized form of conformal welding that will underlie the estimate below.

The next step is to combine this with the boundary equation of motion reviewed in
Section~3. At linear order, the universal JT kernel is
\begin{equation}
B^{(1)}[\delta\theta]
=
\delta\theta''' + \frac{1}{n^2}\delta\theta',
\label{B.12}
\end{equation}
so that, using \eqref{B.5},
\begin{equation}
\partial_\tau B^{(1)}
=
\kappa(n-1)
\left(
\partial_\tau^4+\frac{1}{n^2}\partial_\tau^2
\right)\eta_1
+O\!\left(\kappa (n-1)^2,\kappa^2\right).
\label{B.13}
\end{equation}
Substituting this into the boundary equation of motion from the main text and dividing by
$\kappa(n-1)$, one obtains a linear equation of the schematic form
\begin{equation}
\frac{\phi_r}{4\beta G_N}
\left(
\partial_\tau^4+\frac{1}{n^2}\partial_\tau^2
\right)\eta_1
=
\mathcal W[\eta_1](\tau).
\label{B.14}
\end{equation}
Here $\mathcal W$ denotes the linearized welding-response operator, namely the operator that
encodes how the replicated uniformization data feed into the matter side of the boundary
equation. 

Equation \eqref{B.14} is the only structural fact we will need in what follows. At the first
order relevant for the capacity, the dependence on the replicated saddle enters only through a
linear operator acting on $\eta_1$. The question is therefore how this operator behaves in the
two-QES problem when the factorization parameter is small.

\subsection{Parametric estimate for the first inter-QES correction}

We now turn to the two-QES geometry in the late-time factorized regime. Let $x$ denote the
small parameter controlling the separation of the two replica defects, so that in Lorentzian
kinematics
\begin{equation}
x \sim e^{-2t_0}.
\label{B.15}
\end{equation}
Near each QES one may introduce an adapted local uniformization coordinate. In the limit
$x\to0$, the second defect is pushed to the edge of the corresponding local patch, and the local
geometry reduces to the one-QES problem. This suggests the decomposition
\begin{equation}
\mathcal W
=
\mathcal W_{1\rm QES}^{[1]}
+
\mathcal W_{1\rm QES}^{[2]}
+
\Delta \mathcal W,
\label{B.16}
\end{equation}
where the first two terms are fixed by the two local one-QES building blocks, while
$\Delta\mathcal W$ measures the failure of exact factorization.

The non-trivial issue is the size of $\Delta \mathcal W$. We do not derive it here from an explicit solution
of the full two-QES welding problem. Rather, we use a parametric argument guided by the
exactly solvable $n=2$ case discussed in \cite{HKP2024}. In that case, the first
correction to the factorized off-shell gravitational action begins at order $x^2$. We therefore take
\begin{equation}
\Delta \mathcal W = O(x^2).
\label{deltaW}
\end{equation}

In a patch centered on one QES, the leading
local problem is manifestly the one-QES problem, so the correction produced by the second defect must vanish as $x\to0$.
If the linearized welding operator varies smoothly with the same global uniformization data, its first genuinely inter-QES contribution should begin at that same order. Equation \eqref{deltaW} should therefore be understood as a regularity-based estimate, not as a theorem.

Substituting \eqref{B.16} into \eqref{B.14}, one finds
\begin{equation}
\left[
\frac{\phi_r}{4\beta G_N}
\left(
\partial_\tau^4+\frac{1}{n^2}\partial_\tau^2
\right)
-
\mathcal W_{1\rm QES}^{[1]}
-
\mathcal W_{1\rm QES}^{[2]}
\right]\eta_1
=
\Delta \mathcal W[\eta_1].
\label{B.18}
\end{equation}
Thus the leading solution is the direct sum of two one-QES building blocks, while the first
departure from factorization is sourced by $\Delta\mathcal W$.

If the inverse of the factorized operator on the left-hand side exists and remains regular in the
same regime, then \eqref{deltaW} immediately implies that the first correction to the linearized
solution is also of order $x^2$. It is then natural to write the corresponding capacity as
\begin{equation}
C_I = C_I^{\rm fact} + \Delta C_I^{\rm nf},
\label{B.19}
\end{equation}
where $C_I^{\rm fact}$ is the factorized result computed in Section~4 and $\Delta C_I^{\rm nf}$
is the first genuinely non-factorizing contribution. Under the same regularity assumption,
\begin{equation}
\Delta C_I^{\rm nf}=O(x^2).
\label{B.20}
\end{equation}
Using \eqref{B.15}, this becomes
\begin{equation}
\Delta C_I^{\rm nf}(t_0)=O(e^{-4t_0}).
\label{B.21}
\end{equation}

The role of this appendix is therefore purely parametric. It does not determine the coefficient
of the first non-factorizing correction, nor does it replace a full solution of the global two-QES
welding problem. What it does show is that, once the linearized welding response is assumed to
depend regularly on the global uniformization data, the factorized result of Section~\ref{sec:main} captures
the leading local finite-$n$ effect, while the first omitted inter-QES contribution is additionally
suppressed by $x^2$. This is precisely the sense in which the $O(\kappa^2)$ shift found in the
main text should be understood as the leading local/factorized correction on the island branch.

\section{Including the first omitted mode in the factorized expansion}
\label{app:m3mode}

In the main text we worked in the dominant \(m=2\) truncation of the one-sided replica saddle,
following the factorized large-\(|b|\) regime of \cite{HKP2024}. Since the final one-sided
building block was then evaluated on the representative configuration \(|b|=1\), it is useful to
check explicitly that the first omitted mode, namely \(m=3\), does not alter the leading
\(O(\kappa^2)\) coefficient extracted in Section~\ref{sec:main}. The purpose of this appendix is
precisely to make that point explicit.

The logic is important. The truncation is not justified at \(|b|=1\) by itself. Rather, one first
organizes the one-sided solution in the original large-\(|b|\) factorized regime, where the
boundary deformation is naturally expanded in powers of \(b^{-m}\), and only afterwards evaluates
the resulting truncated building block at \(|b|=1\). The relevant question is therefore whether the
first omitted mode changes the leading large-\(|b|\) coefficient retained in the main text. We now
show that it does not.

The needed linear input from the main text is simply that
\(c_m=-i\kappa\,u_m(n)b^{-m}\), with the leading physical mode \(m=2\) and the first omitted one \(m=3\).
We now examine how the latter corrects the welding maps and the one-sided building block.

\subsection{Subleading effect of the first omitted mode}
\label{appC:welding-m3}

At linear order in the boundary deformation, the corresponding welding maps take the form
\begin{equation}
G(w)
=
w-\gamma(n)\,\frac{\kappa}{b^2}\,w^3
-\eta(n)\,\frac{\kappa}{b^3}\,w^4
+O\!\left(\kappa |b|^{-4} w^5\right),
\label{eq:appC-G}
\end{equation}
and
\begin{equation}
F(z)
=
z-\gamma(n)\,\frac{\kappa}{\bar b^2}\,z^{-1}
-\eta(n)\,\frac{\kappa}{\bar b^3}\,z^{-2}
+O\!\left(\kappa |b|^{-4} z^{-3}\right).
\label{eq:appC-F}
\end{equation}
These formulas extend the leading \(m=2\) truncation used in Section~\ref{sec:welding} by
including the first omitted mode explicitly.

Evaluating the exterior map at the bath endpoint \(z=b\), one obtains
\begin{equation}
F(b)
=
b
-\gamma(n)\,\frac{\kappa}{\bar b^2 b}
-\eta(n)\,\frac{\kappa}{\bar b^3 b^2}
+O\!\left(\kappa |b|^{-7}\right).
\label{eq:appC-Fb-raw}
\end{equation}
For the parametric counting relevant to the factorized regime, this can be written as
\begin{equation}
F(b)
=
b\Bigl(
1-\gamma(n)\,\kappa |b|^{-4}
-\eta(n)\,\kappa |b|^{-6}
+O(\kappa |b|^{-8})
\Bigr).
\label{eq:appC-Fb}
\end{equation}
Similarly,
\begin{equation}
F'(b)
=
1+\gamma(n)\,\kappa |b|^{-4}
+2\eta(n)\,\kappa |b|^{-6}
+O(\kappa |b|^{-8}).
\label{eq:appC-Fpb}
\end{equation}
Thus the \(m=3\) contribution to the endpoint sector is suppressed by an extra factor
\(|b|^{-2}\) relative to the leading \(m=2\) term. This is the first key observation.

The one-sided QES position still begins as
\begin{equation}
a=\frac{\kappa}{\bar b}+O(\kappa^2),
\label{eq:appC-a}
\end{equation}
so \(a=O(\kappa |b|^{-1})\). We may therefore estimate the first omitted contribution to the
interior map evaluated at the QES:
\begin{equation}
\delta G_{(3)}(a)
\sim
\frac{\kappa}{b^3}\,a^4
=
O\!\left(\kappa^5 |b|^{-7}\right).
\label{eq:appC-Ga}
\end{equation}
This is far beyond the \(O(\kappa^2)\) analysis performed in the main text. In particular, the
new mode does not modify the statements
\begin{equation}
G(a)=a+O(\kappa^3),
\qquad
G'(a)=1+O(\kappa^3),
\label{eq:appC-Gaprime}
\end{equation}
which were sufficient for the derivation of \(\alpha_I(n)\) in Section~\ref{sec:main}.

The second key observation is therefore that the first omitted mode only corrects the
\emph{endpoint} data \(F(b)\) and \(F'(b)\), and does so at order \(|b|^{-6}\), while the QES
sector remains unchanged at the order relevant for the \(O(\kappa^2)\) factorized computation.

The point is therefore not simply that the one-sided matter refined entropy is obtained by
truncating the welding data to the $m=2$ mode and then expanding in $\kappa$. Rather, the
coefficient extracted in Section~\ref{sec:main} should be understood as the leading factorized contribution
selected by the retained $m=2$ sector in the original large-$|b|$ hierarchy. The first omitted
mode, $m=3$, does not compete with that contribution: it only enters at the next subleading
order in the same factorized expansion. Equivalently, the finite-$n$ coefficient extracted in
Section~\ref{sec:main} should be viewed as the leading term in the factorized large-$|b|$ hierarchy,
\begin{equation}
\alpha_{mat}(n)
=
\alpha_{mat,2}(n)
+
O(|b|^{-2}),
\label{eq:appC-alpha-hierarchy}
\end{equation}
with
\begin{equation}
\alpha_{mat,2}(n)=8\gamma(n)-\frac12\gamma(n)^2.
\label{eq:appC-alpha-leading}
\end{equation}

We conclude that the first omitted welding mode does not compete with the coefficient
\eqref{eq:appC-alpha-leading}. It only produces the first subleading correction in the same
factorized hierarchy. This is why the \(m=2\) result used in the main text is best interpreted as
the \emph{leading factorized} finite-\(n\) correction in the matter sector, rather than as a result
whose validity depends on setting \(|b|=1\) too early.

\section{Numerical and internal consistency checks}
\label{app:numerics}
In this appendix we provide a numerical validation of the matter coefficient extracted in Section~\ref{sec:main}, together with its replica derivative at $n=1$. All checks are performed within the same factorised $m=2$ truncation used in the analytic calculation. They should therefore be read as consistency tests of the one-sided matter contribution within the same approximation scheme, rather than as probes of the full two-sided replica problem.

\subsection{Numerical extraction of the matter coefficient}
\label{app:numerics-alpha}

The analytic result derived in Section~\ref{sec:main} was
\begin{equation}
\alpha_{mat}(n)=8\gamma(n)-\frac12\gamma(n)^2,
\qquad
\gamma(n)=\frac{3(n^2-1)}{8(4n^2-1)}.
\label{eq:appD-alpha-analytic}
\end{equation}
To test \eqref{eq:appD-alpha-analytic} directly, we start from the reduced one-sided matter
modular entropy at \(|b|=1\),
\begin{equation}
\frac{6n}{c}\,\St_n^{\CFT}
=
\log\frac{(F(b)-a)^2}{(1-a^2)\,F'(b)}
+O(\kappa^3),
\label{eq:appD-reduced-entropy}
\end{equation}
and define the dimensionless quantity
\begin{equation}
y(n,\kappa)\equiv \frac{6n}{c}\,\St_n^{\CFT}.
\label{eq:appD-ydef}
\end{equation}
Within the \(m=2\) truncation, its small-\(\kappa\) expansion has the form
\begin{equation}
y(n,\kappa)=A(n)\kappa+\alpha_{mat}(n)\kappa^2+O(\kappa^3),
\qquad
A(n)=-(2+3\gamma(n)).
\label{eq:appD-small-kappa}
\end{equation}

\begin{table}[!t]
\centering
\begin{tabular}{c c c c c c c c}
\hline
\(n\) & \(A_{\rm fit}\) & \(A_{\rm theory}\) & \(A_{\rm fit}-A_{\rm theory}\) &
\(\alpha_{mat,{\rm fit}}\) & \(\alpha_{mat}^{\rm analytic}\) & difference & \(R^2\) \\
\hline
1.02 & \(-2.01438\) & \(-2.01438\) & \(7.40259\times 10^{-8}\) & \(0.0383118\) & \(0.0383235\) & \(-1.16942\times 10^{-5}\) & \(1.\) \\
1.05 & \(-2.03382\) & \(-2.03382\) & \(7.61080\times 10^{-8}\) & \(0.0901004\) & \(0.0901124\) & \(-1.20296\times 10^{-5}\) & \(1.\) \\
1.10 & \(-2.06152\) & \(-2.06152\) & \(7.96684\times 10^{-8}\) & \(0.1638400\) & \(0.1638520\) & \(-1.26018\times 10^{-5}\) & \(1.\) \\
1.20 & \(-2.10399\) & \(-2.10399\) & \(8.65261\times 10^{-8}\) & \(0.2766960\) & \(0.2767100\) & \(-1.37012\times 10^{-5}\) & \(1.\) \\
1.50 & \(-2.17578\) & \(-2.17578\) & \(1.02163\times 10^{-7}\) & \(0.4670170\) & \(0.4670330\) & \(-1.62015\times 10^{-5}\) & \(1.\) \\
2.00 & \(-2.22500\) & \(-2.22500\) & \(1.15987\times 10^{-7}\) & \(0.5971690\) & \(0.5971880\) & \(-1.84079\times 10^{-5}\) & \(1.\) \\
\hline
\end{tabular}
\caption{Numerical extraction of \(A(n)\) and \(\alpha_{mat}(n)\) from quartic fits in \(\kappa\), using the reduced one-sided expression \eqref{eq:appD-reduced-entropy} at \(|b|=1\) in the factorised \(m=2\) truncation. The fitted coefficient \(\alpha_{mat,{\rm fit}}(n)\) reproduces the analytic result \eqref{eq:appD-alpha-analytic} with high precision throughout the range shown.}
\label{tab:alpha-fit}
\end{table}

For fixed $n$, we evaluate $y(n,\kappa)$ numerically on the grid
\begin{equation}
\kappa \in \{0.006, 0.008, 0.01, 0.012, 0.0125, 0.014, 0.015, 0.016, 0.0175, 0.02, 0.025\},
\end{equation}
and fit the resulting data to a quartic polynomial in $\kappa$,
\begin{equation}
y(n,\kappa)\approx A_{\rm fit}(n)\kappa+\alpha_{mat,\rm fit}(n)\kappa^2+C_3(n)\kappa^3+C_4(n)\kappa^4.
\end{equation}
Here $C_3(n)$ and $C_4(n)$ are merely auxiliary fit coefficients used to absorb higher-order
terms in the small-$\kappa$ expansion; we do not assign them any independent physical meaning.
The fitted coefficient of \(\kappa^2\) can then be compared directly with the analytic prediction
\eqref{eq:appD-alpha-analytic}.

The results are displayed in Table~\ref{tab:alpha-fit}. The agreement is very good throughout the
range shown. The fitted linear coefficient \(A_{\rm fit}(n)\) reproduces the expected value
\(A(n)\) at the level of \(10^{-7}\), while \(\alpha_{mat,{\rm fit}}(n)\) agrees with the analytic
expression \(\alpha_{mat}(n)\) at the level of \(10^{-5}\). The fits are numerically stable, with
\(R^2=1\) to the precision displayed in the table.

\begin{figure}[!t]
\centering
\includegraphics[width=0.78\linewidth]{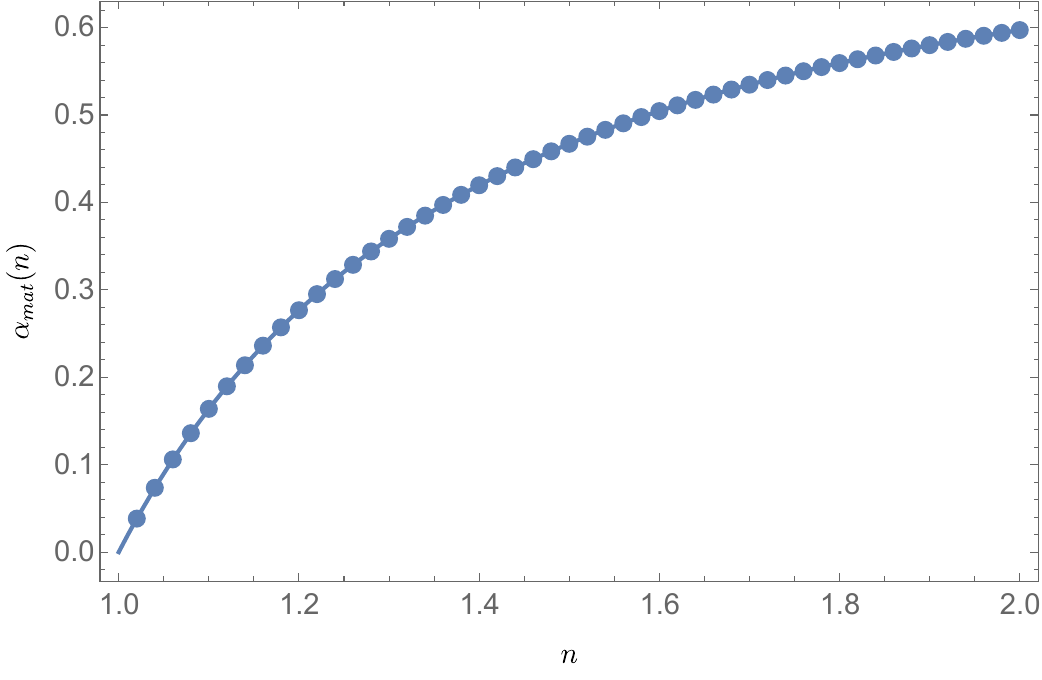}
\caption{Numerical fit check of \(\alpha_{mat}(n)\): fitted points extracted from the quartic fit compared with the analytic expression \eqref{eq:appD-alpha-analytic}. The agreement is excellent throughout the displayed range.}
\label{fig:alpha-fit-check}
\end{figure}

Figure~\ref{fig:alpha-fit-check} provides a visual version of the same check, comparing the fitted
points with the analytic curve \eqref{eq:appD-alpha-analytic}. The agreement is excellent across
the full range considered. This confirms that the matter coefficient $\alpha_{\rm mat}(n)$ obtained in Section~\ref{sec:main} is not an algebraic accident of the intermediate manipulations, but is recovered independently from direct small-$\kappa$ fits of the reduced one-sided matter expression.

\subsection{Replica derivative of the matter coefficient at $n=1$}
\label{app:numerics-derivative}
For the full island-branch capacity, the physically relevant coefficient is the replica derivative of the complete generalized-entropy coefficient discussed in Section~\ref{sec:main}. The purpose of the present subsection is narrower: we test the replica derivative of the matter coefficient extracted in the previous subsection, namely $\alpha_{\rm mat}(n)$, as an independent check of the finite-$n$ structure already present in the one-sided matter contribution. Analytically, Section~\ref{sec:main} gave
\begin{equation}
\alpha_{mat}'(1)=2.
\label{eq:appD-alpha-prime-analytic}
\end{equation}
To test this numerically, we estimate the derivative from the fitted values using a symmetric
finite difference around \(n=1\). This gives
\begin{equation}
\alpha_{mat}'(1)\big|_{\rm fit}=2.00089,
\label{eq:appD-alpha-prime-fit}
\end{equation}
so that
\begin{equation}
\alpha_{mat}'(1)\big|_{\rm fit}-\alpha_{mat}'(1)\big|_{\rm analytic}
=
8.92417\times 10^{-4}.
\label{eq:appD-alpha-prime-diff}
\end{equation}
The discrepancy is well within the level of numerical precision already visible in Table~\ref{tab:alpha-fit}. In particular, the fit confirms that the one-sided matter coefficient has a nonzero and numerically stable replica derivative at $n=1$, in agreement with the analytic value $2$.

This check should be interpreted with some care. The quantity analyzed here is the matter coefficient $\alpha_{\rm mat}(n)$, not the full island coefficient that enters the generalized modular entropy after the on-shell dilaton contribution is included. What the numerical result shows is that the matter sector by itself already carries a nontrivial nearby finite-$n$ structure. The physical capacity shift discussed in the main text is obtained only after combining this matter contribution with the analytic dilaton correction derived in Section~\ref{sec:main}.

\end{document}